\documentclass[letterpaper,12pt]{article}

\usepackage{amssymb}
\usepackage{amsbsy}

\newbox\stroke\setbox\stroke\hbox{}                       
\newdimen\spac\spac=.5mm
\newdimen\xm\xm=90mm
\newdimen\xcntr\xcntr=45mm
\newcount\multipler\multipler=1
\def\dott{\vrule height .5mm depth 0mm}
\def\fill{
\setbox\stroke=\hbox{\unhbox\stroke\hskip\spac\dott}
\multiply\spac by \multipler\divide\spac by 1000
\ifdim\wd\stroke>\xcntr\multipler=950\fi
\ifdim\wd\stroke<\xcntr\multipler=1050\fi
\ifdim\wd\stroke<\xm\expandafter\fill\fi} \fill
\def\barrier{\copy\stroke}

\newcommand{\ie}{\emph{i.e.}}                             
\newcommand{\eg}{\emph{e.g.}}                             
\newcommand{\cf}{\emph{cf.}}                              
\newcommand{\etc}{\emph{etc.}}                            
\newcommand{\alev}{\textrm{a.e.}}                         
\newcommand{\rhs}{\emph{r.h.s.}}                          
\newcommand{\qed}
           {\mbox{\quad\rule[-1.5pt]{.4em}{1.5ex}}}       

\newcommand{\Real}{\mathop{\mathbb R}\nolimits}           
\newcommand{\Com}{\mathop{\mathbb C}\nolimits}            
\newcommand{\Nat}{\mathop{\mathbb N}\nolimits}            
\newcommand{\supp}{\mathop{\mathrm{supp}}\nolimits}       
\newcommand{\sgn}{\mathop{\mathrm{sgn}}\nolimits}         
\newcommand{\re}{\mathop{\mathrm{Re}}\nolimits}           
\newcommand{\im}{\mathop{\mathrm{Im}}\nolimits}           
\newcommand{\Ker}{\mathop{\mathrm{Ker}}\nolimits}         
\newcommand{\tr}{\mathop{\mathrm{tr}}\nolimits}           

\newcommand{\si}{\boldsymbol{L}^1}                        
\newcommand{\sii}{\boldsymbol{L}^2}                       
\newcommand{\sieps}{\boldsymbol{L}^{1+\varepsilon}}       
\newcommand{\sobi}{\boldsymbol{W}_{\!0}^{2,1}}            
\newcommand{\Comp}{\boldsymbol{C}_{\!0}^\infty}             
\newcommand{\Smooth}{\boldsymbol{C}^\infty}               

\newcommand{\PF}{\textsc{Proof:\quad}}                    
\newtheorem{claim}{Claim}[section]
\newtheorem{prop}[claim]{Proposition}                     
\newtheorem{thm}[claim]{Theorem}                          
\newtheorem{corol}[claim]{Corollary}                      
\newtheorem{lemma}[claim]{Lemma}                          
\newtheorem{rems}[claim]{Remarks}                         

\newcommand{\beq}{\begin{equation}}                       
\newcommand{\eeq}{\end{equation}}                         
\newcommand{\beqn}{\begin{eqnarray}}                      
\newcommand{\eeqn}{\end{eqnarray}}                        

\def\OMIT#1{}                                             

\begin{document}

\noindent
\begin{flushleft}
{\large WAVEGUIDES COUPLED THROUGH \\[0.5mm]  A SEMITRANSPARENT
BARRIER: \\[1.5mm] A BIRMAN-SCHWINGER ANALYSIS }
\\ \rule{137truemm}{1pt}
\end{flushleft}
\vspace{5mm}
\begin{flushright}
P. EXNER \\ [1mm] Department of Theoretical Physics, NPI
\\ Academy of Sciences, 25068 \v Re\v z near Prague \\
and Doppler Institute, Czech Technical University \\
B\v{r}ehov{\'a} 7, 11519 Prague, Czechia \\ {\em exner@ujf.cas.cz}
\vspace{5mm}

D. KREJ\v{C}I\v{R}\'{I}K \\ [1mm] Department of Theoretical
Physics Institute, NPI AS, \\ Faculty of Mathematics and Physics,
Charles University \\ V~Hole\v{s}ovi\v{c}k\'ach~2, 18000 Prague,
Czechia, \\
and Facult\'{e} des Sciences et Technologies \\
Universit\'{e} de Toulon et du Var \\ BP~132, 83957 La Garde
Cedex, France \\  {\em krejcirik@ujf.cas.cz}
\end{flushright}
\vspace{5mm}

\noindent {\small The paper is devoted to a model of a mesoscopic
system consisting of a pair of parallel planar waveguides
separated by an infinitely thin semitransparent boundary modeled
by a transverse $\delta$ interaction. We develop the
Birman-Schwinger theory for the corresponding generalized
Schr\"odinger operator. The spectral properties become nontrivial
if the barrier coupling is not invariant with respect to
longitudinal translations, in particular, there are bound states
if the barrier is locally more transparent in the mean and the
coupling parameter reaches the same asymptotic value in both
directions along the guide axis. We derive the weak-coupling
expansion of the ground-state eigenvalue for the cases when the
perturbation is small in the supremum and the $\si$-norms. The
last named result applies to the situation when the support of the
leaky part shrinks: the obtained asymptotics differs from that of
a double guide divided by a pierced Dirichlet barrier. We also
derive an upper bound on the number of bound states . }

\vspace{5mm}

\tableofcontents

\setcounter{equation}{0}
\section{Introduction}
\subsection{Motivation}
The recent progress of solid-state physics opened way to testing
of quantum mechanics in hitherto unusual situations. Many of the
``mesoscopic" semiconductor systems can be regarded as electron
waveguides in which wave properties of the particles play an
essential role -- we refer to \cite{DE1, ESTV} for discussion of
the model assumptions involved and a bibliography.

An interesting class of such systems is represented by a pair of
parallel planar guides with a lateral coupling, which is realized
either by a ``window" in a Dirichlet barrier separating the ducts
\cite{BGRS, ESTV} or by a local variation of the coupling
parameter in a leaky, \ie, semitransparent barrier \cite{EK1}. In
the latter case (sketched in Figure~1) the Hamiltonian is formally
given by the relation (\ref{formal}) below, with the barrier
supported by the $x$--axis. The function $\alpha$ describes the
coupling parameter and the outer boundary of the double strip
$\Omega:=\Real\times(-d_2,d_1)$ is supposed to be hard, \ie, we
impose Dirichlet boundary conditions there.

\begin{figure}
{\fontsize{10pt}{0}
\setlength{\unitlength}{1mm}
\begin{picture}(120,30)(-12,10)
  \put(10,14){\rule{\xm}{1mm}}
  \put(10,30){\rule{\xm}{1mm}}
  \put(10,19.8){\barrier}
  \put(0,20){\line(1,0){12}}
  \put(100,20){\vector(1,0){10}}
  \put(55,8){\vector(0,1){30}}
  \put(110,17){$x$}
  \put(57,38){$y$}
  \put(48.5,10.1){$-d_{2}$}
  \put(51.5,32.5){$d_{1}$}
\end{picture}
\caption{Double waveguide with a $\delta$ barrier}\label{schm}
}
\end{figure}
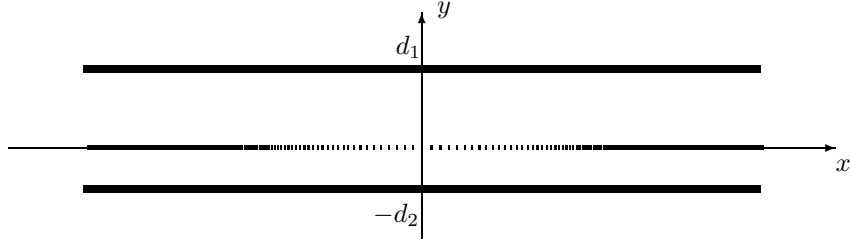

Depending on the choice of $\alpha$, such a model describe a
variety of different dynamical situations. It is illustrative to
consider the case related to the example of a pierced-hard-wall
discussed in Ref.~\cite{ESTV}; the comparison being based on the
fact that the $\delta$ interaction with a large coupling constant
approximates the Dirichlet barrier. The example of a
step-function-shaped $\alpha$ analyzed in Refs.~\cite{EK1}
exhibits indeed for large $\alpha$ close similarities in the
numerically calculated shapes of the eigenfunctions, \etc$\;$ At
the same time, asymptotic properties of the eigenvalues may be
rather different in the two cases if we exclude here the
possibility $\alpha=\infty$ which expresses formally the Dirichlet
boundary condition -- \cf~Remarks~\ref{formallimit}.

Systems with a $\delta$ potential barrier of the type
(\ref{formal}) are mathematically more accessible, since two
operators with different functions $\alpha$ have the same form
domain. This observation will make it possible to construct a
Birman-Schwinger-type theory in this case writing down an explicit
expression for the difference between the resolvent of the
Hamiltonian (\ref{formal}) and that of a suitable comparison
operator. The main consequence of this formula which we derive in
this paper is the weak coupling expansion in the situation when
$\alpha$ forms a ``potential well", \ie, when the barrier is
locally more transparent (at least in the mean) and the coupling
parameter $\alpha(x)$ reaches the same asymptotic value in both
directions along the guide axis.

\subsection{Description of the model and contents of the paper}
\label{model}
As we have said, the configuration space of our system is a
straight planar strip $\Omega:=\Real\times\mathcal{O}$ with
$\mathcal{O}:=\mathcal{O}_2\cup\mathcal{O}_1:=(-d_2,0)\cup(0,d_1)$
in which the free motion is restricted by the outer hard walls and
a $\delta$ potential barrier at $y=0$. Its coupling strength
$\alpha\in\Real$ varies longitudinally, $\alpha=\alpha(x)$, so the
particle Hamiltonian can be formally written as
\begin{equation} \label{formal}
H_{\alpha}= -\Delta^{\Omega}_D+\alpha(x)\delta(y)\,.
\end{equation}
There are several equivalent ways to give the right-hand-side of
(\ref{formal}) a rigorous meaning. Following \cite[Chap.~I.3]{AGH}
this can be done by imposing the standard boundary
conditions~\cite{EK1}. In this paper, however, we use instead a
quadratic-form definition which is much more general. Such
generalized Schr\"odinger operators in $\Real^d$ with a
measure-induced interaction were studied in \cite{BEKS}. In the
next section we shall adapt this theory for the case when the free
Hamiltonian is the Dirichlet Laplacian relative to a subset
$\Omega\subset\Real^d$.~To make the paper self-contained, we
outline the construction from Sec.~2 of the mentioned paper with
emphasis on the modifications required by the presence of the
Dirichlet boundary. This concerns mostly Lemma~\ref{lem2.2} whose
proof in \cite{BEKS} relies on the explicit form of the free
Green's function.

In Section~3 we shall use this results to formulate the
Birman-Schwinger theory for the operator (\ref{formal}). The basic
idea is again adopted from \cite{BEKS}, however, it suits to our
purpose to express the resolvent difference with respect to a
comparison operator which also has a nonzero $\alpha$, and to
write it in a symmetric form.

The obtained resolvent formula is then employed to investigate the
discrete spectrum of our Hamiltonian which exists if the $\delta$
barrier produces a local attractive interaction. Sections~4 and 5
are devoted the weak-coupling analysis of our model in two
situations: in the first case the interaction is tuned by means of
a coupling constant, in the second one we use instead a scaling
transformation with respect to the longitudinal variable. In both
situations we derive an asymptotic expansion for the ground state
eigenvalue. In the example which involves the scaling the
``potential well" given by the shape of the function $\alpha$ may
be deep, the weak coupling being achieved by its narrowness. This
makes it possible to compare the asymptotics with the mentioned
Dirichlet case, where the gap is proportional to the fourth power
of the window width. In contrast, for any ``soft" barrier the
width appears in the leading term of the asymptotics with the
square only.

In the final section we use the Birman-Schwinger technique to
derive an upper bound on the dimension of the discrete s pectrum.
A comparison with the ``square well" example of~\cite{EK1} shows
that the bound is good for weak coupling but its semiclassical
behavior is not correct as it is the case for the usual
Schr\"odinger operators \cite{Ne}.

\setcounter{equation}{0}
\section{Singularly supported interactions on \\ a subset of~$\Real^{d}$}
\label{Sec.BS}

Let $\Omega$ be an open subset of~$\Real^{d}$. Consider a positive
Radon measure $m$ on $\Omega$, \ie\/ the abstraction of Lebesgue's
outer measure for general topological spaces
\cite[Def.~2.3.9]{Rao}, and a Borel measurable function
$\alpha:\Real^d\rightarrow\Real$ such that
\begin{equation}\label{bass}
  \int_{\Omega}|\psi(x)|^2 \left(1+\alpha(x)^2\right) dm(x)\leq
  a\int_{\Omega}|\nabla\psi(x)|^2 dx+b\int_{\Omega}|\psi(x)|^2 dx
\end{equation}
holds for all $\psi\in\Comp(\Omega)$ and some positive $a<1$ and
$b$. As indicated in the introduction, we are interested mainly in
situation when $m$ is a $\delta$-measure supported by a planar
curve, but the argument presented below does not need such
restrictions on the measure or space dimension; it includes also
the regular potential case, $dm(x)=|V(x)|dx$.

By definition, $\Comp(\Omega)$ is dense in the local Sobolev
space $\sobi(\Omega)$ (\cf~\cite[Sec.~XIII.14]{RS}), so there is
a unique bounded linear operator
\begin{displaymath}
  I_{m}:\sobi(\Omega)\rightarrow\sii(m):=\sii(\Omega,dm)
\end{displaymath}
such that $I_{m}\psi=\psi$ is valid for any
$\psi\in\Comp(\Omega)$. The last relation means in fact
$(I_{m}\psi)(x)=\psi(x)$ for $x\in\supp m$; with an abuse of
notation we shall employ the symbol $\psi$ for $(i)$ a continuous
function $\psi$, $\;(ii)$ the corresponding $\sii(\Omega)$
equivalence class, and finally $(iii)$ for the corresponding
$\sii(\Omega,dm)$ equivalence class. By density, the
inequality~(\ref{bass}) holds for all $\psi\in\sobi(\Omega)$
provided $\psi$ is replaced by $I_{m}\psi$ on the left-hand-side.

\subsection{The Hamiltonian}\label{Sec.BS1}

We introduce the following quadratic form
\begin{equation}\label{formE}
  {\cal E}_{\alpha m}(\psi,\varphi) :=
  \int_{\Omega}\overline{\nabla\psi(x)}.\nabla\varphi(x) dx+
  \int_{\Omega}\alpha(x)\, (I_{m}\bar\psi)(x)\, (I_{m}\varphi)(x)\,
   dm(x)
\end{equation}
with the domain $D({\cal E}_{\alpha m})= \sobi(\Omega)$. It is
well known \cite[Sec.~XIII.14]{RS} that the free form ${\cal
E}_{0}$ is positive and closed on $\sii(\Omega)$; it gives rise to
the Dirichlet Laplacian $-\Delta_{D}^{\Omega}$. By the KLMN
theorem \cite[Thm.~X.17]{RS} and the extended version of
inequality~(\ref{bass}), ${\cal E}_{\alpha m}$ is lower
semibounded and closed on $\sii(\Omega)$, and $\Comp(\Omega)$ is a
core for it. Hence by the second representation theorem, there is
a unique self-adjoint operator $H_{\alpha m}$ associated with
${\cal E}_{\alpha m}$; it will the object of our interest in the
following.

The basic assumption (\ref{bass}) is satisfied, in particular, for
measures $m$ belonging to the generalized Kato class. By~\cite{SV}
the inequality (\ref{bass}) holds for such $m$ and $\psi\in{\cal
S}(\Real^{d})$, and the same is, {\em a fortiori}, true for
$\psi\in\Comp(\Omega)$ corresponding to an open
$\Omega\subset\Real^{d}$. If $d=2$ the Kato condition reads
\begin{equation}
  \lim_{\varepsilon\to 0+}\sup_{x\in\Omega}\int_{B(x,\varepsilon)\cap\Omega}
  \left|\ln|x-y|\right| dm(y)=0\,,
\end{equation}
where $B(x,\varepsilon)$ is the ball of radius $\varepsilon$ and
center $x$. It is straightforward to check that the condition is
satisfied for the $\delta$ measure of our example; alternatively
one can employ Thm.~4.1 of~\cite{BEKS}.

\subsection{Auxiliary results}

We will need two lemmas. The first one is abstract and we adopt it
from \cite{BEKS}:
\begin{lemma}\label{lem2.1}
Let ${\cal E}$ be a lower semibounded densely defined closed
quadratic form on a complex Hilbert space ${\cal H}$ with the
inner product $(\cdot\, ,\cdot)$, and let $H$ be the unique
self-adjoint operator on ${\cal H}$ associated with ${\cal E}$.
Finally, let $R:{\cal H}\rightarrow D({\cal E})$ be an arbitrary
map and $z\in\Com$. Then the following statements are equivalent.
\begin{eqnarray*}
  &\mbox{\rm (i)} & z\in\rho(H)\ \mbox{\it and}\ \ (H-z)^{-1}=R.\\
  &\mbox{\rm (ii)}& \forall\,\psi\in{\cal H}, \varphi\in D({\cal E})\ :\quad
         {\cal E}(R\psi,\varphi)=(z R\psi+\psi,\varphi).
\end{eqnarray*}
\end{lemma}
\vspace{3mm}

Let now $\Com_{\Omega,0}^+$ be the set $\left\{k: \: \im k>0
\,\;\mathrm{or}\; k^2\in \left[0,\inf\sigma(-\Delta^{\Omega}_D) \right)
\right\}\subset \Com$. Given $k\in\Com_{\Omega,0}^+$ we denote by
$G_{0}(\cdot,\cdot;k)$ the free resolvent kernel for $z=k^2$
corresponding to the Dirichlet Laplacian $-\Delta_{D}^{\Omega}$.
The main difference with respect to~\cite{BEKS} is that for
$\Omega\not=\Real^{d}$ the kernel depends on both arguments, not
just on their difference.

Let $\mu,\nu$ be positive Radon measures without a discrete
component, \ie\, $\mu(\{a\})=\nu(\{a\})=0$ for any $a\in\Omega$.
We denote by $R_{\mu,\nu}^{k}$ the integral operator from
$\sii(\mu):=\sii(\Omega,d\mu)$ to $\sii(\nu)$ with the kernel
$G_{0}(\cdot,\cdot;k)$. In particular, we have
\begin{equation}
  (R_{\mu,\nu}^{k}\psi)(x)=\int_{\Omega}G_{0}(x,y;k)\psi(y) d\mu(y)
\end{equation}
for all $\psi\in D(R_{\mu,\nu}^{k})\subset\sii(\mu)$. Since we
agreed to the mentioned abuse of notation, there is no $\nu$ in
the definition; we compute the right-hand-side and interpret it as
values of a function in $\sii(\nu)$. In the following $(\cdot\, ,\cdot)$ will
denote the inner product on $\sii(\Omega)$.

\begin{lemma}\label{lem2.2}
  Let $k\in\Com_{\Omega,0}^+$ and $\psi\in\sii(m)$.
Then $R_{m,dx}^{k}\psi\in\sobi(\Omega)$ and
$$
  \forall\varphi\in\sobi(\Omega):\
  \quad{\cal E}_{0}(R_{m,dx}^{k}\psi,\varphi)-(k^2 R_{m,dx}^{k}\psi,\varphi)=
   \int_{\Omega}\bar\psi(y)(I_{m}\varphi)(y) dm(y)
$$
In particular, $R_{m,dx}^{k}$ is injective.
\end{lemma}
\PF If we prove the above relation, the injectivity will follow by
density of Ran $I_{m}$ in $\sii(m)$. Assume first $k^2<0$, \ie,
$k$ is purely imaginary. Then
\begin{equation}
  \left<\psi,\varphi\right>_{k}:={\cal E}_{0}(\psi,\varphi)
                                 -k^2(\psi,\varphi)
\end{equation}
defines an inner product on $\sobi(\Omega)$ and the corresponding
norm is equivalent to the usual Sobolev norm (with $k^2\!=\!-1$).
Take a fixed $\psi\in\sii(m)$. Using Schwarz inequality and the
fact that $I_{m}$ is bounded we infer
\begin{eqnarray*}
  \left|\int_{\Omega}\bar\psi(y)(I_{m}\varphi)(y)dm(y)\right|^2
&\leq&
  \int_{\Omega}|\psi(y)|^2 dm(y)\int_{\Omega}|(I_{m}\varphi)(y)|^2 dm(y)\\
&\leq&
  c\left<\varphi,\varphi\right>_{k}
\end{eqnarray*}
for any $\varphi\in\sobi(\Omega)$ and some constant $c$ depending
on $\psi$. Hence the linear functional
\begin{displaymath}
  \varphi\mapsto\int_{\Omega}\bar\psi(y)(I_{m}\varphi)(y)dm(y)
\end{displaymath}
on the Hilbert space $(\sobi(\Omega),\left<\cdot\, ,\cdot\right>_{k})$
is bounded, and by Riesz's lemma, there is a unique
$\psi_{m}^{k}\in\sobi(\Omega)$ such that
\begin{equation}\label{2.3}
  \forall\varphi\in\sobi(\Omega):\quad
  \left<\psi_{m}^{k},\varphi\right>_{k}=
  \int_{\Omega}\bar\psi(y)(I_{m}\varphi)(y)dm(y).
\end{equation}
Consequently, it is sufficient to show that
\begin{displaymath}
  \forall\psi\in\sii(\Omega):\quad
  (R_{m,dx}^{k}\psi)(x)\equiv\int_{\Omega}G_{0}(x,y;k)\psi(y)dm(y)
  =\psi_{m}^{k}(x)
\end{displaymath}
a.e. with respect to the Lebesgue measure $dx$.

If $\Omega\ne\Real^d$ we have in general no explicit expression
for the Green's function $G_{0}(x,y;k)$ with a given $k^2<0$. We
know, however, that it is positive for all $x,y\in\Omega,
x\not=y\;$ \cite[App.~1 to Sec.~XIII.12]{RS}, and moreover, that
the kernel is $dx$-integrable if the other variable is fixed (in
fact, exponentially decaying for a non-compact $\Omega$) and
\begin{displaymath}
  \forall\,y\in\Omega,\varphi\in\sobi(\Omega):\quad
  \int_{\Omega}G_{0}(x,y;k)(-\Delta_{D}^{\Omega}-k^2)\varphi(x) dx
  =\varphi(y).
\end{displaymath}
Functions $\varphi:= (-\Delta_{D}^{\Omega}-k^2)^{-1}\eta$ with
$\eta\in\Comp(\Omega)$ are bounded $\Smooth$ and have the same
decay as the Green's function for a non-compact $\Omega$.

We shall prove the desired relation in several steps. Suppose
first that $\psi\in\si(m)\cap\sii(m)$. For $\varphi$ of the
described class we may employ then the Fubini theorem obtaining
\begin{eqnarray*}
\lefteqn{\int_{\Omega}\overline{\left(\int_{\Omega}G_{0}(x,y;k)
  \psi(y)dm(y)\right)}
               (-\Delta_{D}^{\Omega}-k^2)\varphi(x)\,dx} \\
&&\qquad=\int_{\Omega}\bar\psi(y)
                \left(\int_{\Omega}G_{0}(y,x;k)
                (-\Delta_{D}^{\Omega}-k^2)\varphi(x)\,dx\right)dm(y)\\
&&\qquad=\int_{\Omega}\bar\psi(y)\varphi(y)\,dm(y)\,.
\end{eqnarray*}
We have used here also the fact that $G_{0}$ is real-valued for
$k^2<0$. By~(\ref{2.3}) and the second representation theorem, we
have
$$
  \int_{\Omega}\overline{\psi_{m}^{k}}
  (-\Delta_{D}^{\Omega}-k^2)\varphi(x)\,dx
  =\left<\psi_{m}^{k},\varphi\right>_{k}
  =\int_{\Omega}\bar\psi(y)\varphi(y)\,dm(y)
$$
for all $\varphi\in (-\Delta_{D}^{\Omega}-k^2)^{-1}\Comp(\Omega)$.

In the last equality, we have used the fact that
$I_m\varphi=\varphi$~~$m-\alev$ This relation is not selfevident
because in general~$\varphi$ does not belong to~$\Comp(\Omega)$.
However, one can approximate it by~$\Comp$--functions. More
specifically, define~$\varphi_n:=j_n \varphi$,
where~$j_n\in\Comp(\Omega)$ such that~$0\leq j_n(x) \leq 1$ for
all~$x\in\Omega$ and~$j_n(x)=1$ for~$|x|\leq n$. Since
$\varphi_n\to\varphi$ pointwise~$dx-\alev$ as~$n\to\infty$ and
$|\varphi_n|\leq|\varphi|\in\sobi(\Omega)$, it follows by the
dominated convergence theorem that~$\varphi_n\to\varphi$
in~$\sobi(\Omega)$. From the definition of~$I_m$, we get
also~$I_m\varphi_n\to I_m\varphi$ in~$\sii(m)$.
Since~$I_m\varphi_n\in\Comp(\Omega)$ by construction, we infer
that~$I_m\varphi_n=\varphi_n$ holds~$m-\alev$, and
therefore~$I_m\varphi=\varphi$~~$m-\alev$ as well.

Since the set~$(-\Delta_{D}^{\Omega}-k^2)^{-1}\Comp(\Omega)$
is dense in $\si(\Omega)$, it follows that
\begin{displaymath}
  \psi_{m}^{k}=R_{m,dx}^{k}\psi \qquad dx\!-\!\alev
\end{displaymath}
Now we can mimick the argument of \cite{BEKS} again: in the next
step we consider a non-negative $\psi\in\sii(m)$ and use a
standard approximation argument choosing a sequence
$\{\tilde\psi_{n}\}_{n=1}^{\infty} \subset\si(m)\cap\sii(m)$ such
that
$$ \lim_{n\to\infty}\tilde\psi_{n}=\psi \quad {\rm and} \quad
0\leq\tilde\psi_{1}\leq\tilde\psi_{2}\leq\cdots\leq\tilde\psi_{n}
\qquad m\!-\!\alev $$
Then
\begin{displaymath}
  \forall\varphi\in\sobi(\Omega):\quad
  \left<R_{m,dx}^{k}\tilde\psi_{n},\varphi\right>_{k}
  =\int_{\Omega}\overline{\tilde\psi_{n}(y)}(I_{m}\varphi)(y)\,dm(y)
\end{displaymath}
and by the dominated convergence theorem we get
\begin{displaymath}
  \forall\varphi\in\sobi(\Omega):\quad
  \left<R_{m,dx}^{k}\tilde\psi_{n},\varphi\right>_{k}
  \rightarrow\int_{\Omega}\bar\psi(y)(I_{m}\varphi)(y)\,dm(y)
  =\left<R_{m,dx}^{k}\psi,\varphi\right>_{k}
\end{displaymath}
as $n\to\infty$, \ie, the sequence
$\{R_{m,dx}^{k}\tilde\psi_{n}\}$ converges weakly to
$R_{m,dx}^{k}\psi$ in the Hilbert space
$(\sobi(\Omega),\left<\cdot\, ,\cdot\right>_{k})$. By the diagonal
trick \cite[Sec.~I.5]{RS} we may thus assume (selecting a
subsequence if necessary) that
\begin{equation}\label{2.5}
  R_{m,dx}^{k}\psi_{n}\rightarrow R_{m,dx}^{k}\psi
  \quad\mbox{as}\quad n\to\infty
\end{equation}
strongly in $\sobi(\Omega)$, where
$\psi_{n}:=\frac{1}{n}\sum_{j=1}^{n}\tilde\psi_{j}$.
Since $G_{0}(\cdot,\cdot;k)$ is nonnegative and the sequence $\{\psi_{n}\}$
is nondecreasing again, the monotone convergence theorem implies
\begin{displaymath}
  \forall x\in\Omega:\quad
  \int_{\Omega}G_{0}(x,y;k)\psi(y)\,dm(y)
  =\lim_{n\to\infty}(R_{m,dx}^{k}\psi_{n})(x)\leq\infty.
\end{displaymath}
The relation~(\ref{2.5}) then gives
\begin{displaymath}
  \int_{\Omega}G_{0}(\cdot,y;k)\psi(y)\,dm(y)=R_{m,dx}^{k}\psi
  \qquad dx\!-\!\alev
\end{displaymath}
Finally, by linearity the result extends to any $\psi\in\sii(m)$.

It remains to establish the sought identity for an arbitrary
$k\in\Com_{\Omega,0}^+$. To this aim, we employ the first
resolvent
relation which gives
\begin{displaymath}
  R_{m,dx}^{\tilde k}=
  R_{m,dx}^{k}+(\tilde k^2-k^2)R_{0}^{\tilde k}R_{m,dx}^{k},
\end{displaymath}
where we have denoted $R_{0}^{k}:=R_{dx,dx}^{k}$. Using repeatedly
Lemma~\ref{lem2.1} we find
\begin{eqnarray*}
&&{\cal E}_{0}(R_{m,dx}^{\tilde k}\psi,\varphi)
  -(\tilde k^2 R_{m,dx}^{\tilde k}\psi,\varphi)\\
\\
&&\qquad
  ={\cal E}_{0}(R_{m,dx}^{k}\psi,\varphi)
  -(\tilde k^2 R_{m,dx}^{k}\psi,\varphi)\\
&&\qquad
  \quad
  +{\cal E}_{0}((\tilde k^2-k^2)R_{0}^{\tilde k}R_{m,dx}^{k}\psi,\varphi)
  -(\tilde k^2(\tilde k^2-k^2)R_{0}^{\tilde k}R_{m,dx}^{k}\psi,\varphi)\\
\\
&&\qquad
  ={\cal E}_{0}(R_{m,dx}^{k}\psi,\varphi)
  -(\tilde k^2 R_{m,dx}^{k}\psi,\varphi)
  +(\tilde k^2(\tilde k^2-k^2)R_{0}^{\tilde k}R_{m,dx}^{k}\psi,\varphi)\\
&&\qquad
  \quad
  +((\tilde k^2-k^2)R_{m,dx}^{k}\psi,\varphi)
  -(\tilde k^2(\tilde k^2-k^2)R_{0}^{\tilde
  k}R_{m,dx}^{k}\psi,\varphi)\\
\\
&&\qquad
  ={\cal E}_{0}(R_{m,dx}^{k}\psi,\varphi)
  -(k^2 R_{m,dx}^{k}\psi,\varphi).\qed
\end{eqnarray*}

\subsection{The resolvent}
The above result allows us to write an explicit formula for the
resolvent of $H_{\alpha m}$ and to derive some properties of it.
We could just quote the results which we shall need in the
following, but for the sake of completeness we sketch also the
proofs which are essentially the same as in \cite{BEKS}.
\begin{prop} \label{resolv1}
Let $k\in\Com_{\Omega,0}^+$. Suppose that the operator $I+\alpha
I_{m}R_{m,dx}^{k}$ is invertible on $\sii(m)$ and the operator %
\begin{displaymath}
  R^{k}:=R_{0}^{k}-R_{m,dx}^{k}(I+\alpha I_{m}R_{m,dx}^{k})^{-1}
                   \alpha I_{m}R_{0}^{k}
\end{displaymath}
is defined everywhere in $\sii(\Omega)$. Then
$k^2\in\rho(H_{\alpha m})$ and $(H_{\alpha m}-k^2)^{-1}=R^{k}$.
\end{prop}
\PF Take $\psi\in\sii(\Omega)$ and $\varphi\in\sobi(\Omega) \equiv
D({\cal E}_{\alpha m})$. By assumption, the operator $R^{k}$ is
defined on $\sii(\Omega)$. The free resolvent maps $\sii(\Omega)$
into $\sobi(\Omega)$; the same is true for the second term in view
of the assumed invertibility and Lemma~\ref{lem2.2}. Thus
$R^{k}\psi\in\sobi(\Omega)$, and by Lemma~\ref{lem2.1} we have to
check that
\begin{displaymath}
  {\cal E}_{\alpha m}(R^{k}\psi,\varphi)-(k^2 R^{k}\psi,\varphi)
  =(\psi,\varphi)
\end{displaymath}
holds for all $\psi,\varphi$ from the indicated sets. Dividing the
left-hand-side into the ``free'' and ``interaction'' parts and
denoting
\begin{displaymath}
  \chi:=(I+\alpha I_{m}R_{m,dx}^{k})^{-1}\alpha I_{m}R_{0}^{k}\psi,
\end{displaymath}
we can rewrite it as
\begin{eqnarray*}
\lefteqn{{\cal E}_{0}(R_{0}^{k}\psi,\varphi)-(k^2
R_{0}^{k}\psi,\varphi)-{\cal E}_{0}(R_{m,dx}^{k}\chi,\varphi)+(k^2
R_{m,dx}^{k}\chi,\varphi) }\\ && \qquad
+\int_{\Omega}\alpha(x)(I_{m}\overline{R^{k}\psi})(x)
(I_{m}\varphi)(x)\,dm(x) \phantom{AAAA}
\end{eqnarray*}
The first two pair of terms equal $(\psi,\varphi)$ and
$-\int_{\Omega}\bar\chi(x)(I_{m}\varphi)(x)\,dm(x)$  by
Lemma~\ref{lem2.1} and Lemma~\ref{lem2.2}, respectively. Since the
relation should hold for all $\varphi\in\sobi(\Omega)$ we have
thus to check that $\alpha I_{m}R^{k}\psi=\chi$ for any
$\psi\in\sii(\Omega)$, which follows by a simple algebraic
manipulation,
\begin{eqnarray*}
\lefteqn{\alpha I_{m}R^{k}\psi =\alpha I_{m}R_{0}^{k}\psi-\alpha
I_{m}R_{m,dx}^{k}\chi} \\ && = (I+\alpha I_{m}R_{m,dx}^{k}\psi)
    \underbrace{(I+\alpha I_{m}R_{m,dx}^{k}\psi)^{-1}
    \alpha I_{m}R_{0}^{k}\psi}_{\chi}
 -\alpha I_{m}R_{m,dx}^{k}\chi=\chi\,. \;\qed
\end{eqnarray*}

As usual the invertibility assumption of the preceding proposition
is satisfied for energies large enough negative.
\begin{corol}\label{Cor.2.4}
There is $\kappa_{0}>0$ such that $\|\alpha I_{m}
R_{m,dx}^{i\kappa}\|<1$ for $\kappa\ge\kappa_0$.
\end{corol}
\PF In view of our basic assumption (\ref{bass}) and the
boundedness of the operator $I_{m}$ we can choose $a<1$ and
$0<b<\infty$ such that
$$ \forall\varphi\in\sobi(\Omega):\quad
  \int_{\Omega}|I_{m}\varphi(x)|^2 (1+\alpha(x)^2)\,dm(x)
  \leq a\left<\varphi,\varphi\right>_{i\kappa_{0}}
$$
where $\kappa_{0}:=\sqrt{\frac{b}{a}}$ and the inner product at
the \rhs\/ is defined in the proof of Lemma~\ref{lem2.2}. We
denote by $S_{\kappa}$ the unit sphere in
$(\sobi(\Omega),\left<\cdot\, ,\cdot\right>_{i\kappa})$. Given
$\kappa\geq\kappa_{0}$ and $\psi\in\sii(m)$, we deduce from the
above inequality
\begin{eqnarray*}
\lefteqn{\int_{\Omega}\alpha(x)^2 |(I_{m}R_{m,dx}^{i\kappa}\psi)(x)|^2 dm(x)
\leq a\left<R_{m,dx}^{i\kappa}\psi,
              R_{m,dx}^{i\kappa}\psi\right>_{i\kappa_{0}}}\\
&&\qquad\qquad \leq a\left<R_{m,dx}^{i\kappa}\psi,
              R_{m,dx}^{i\kappa}\psi\right>_{i\kappa}
=a\sup_{\varphi\in S_{\kappa}}
       \left|\left<R_{m,dx}^{i\kappa}\psi,\varphi\right>_{i\kappa}\right|^2.
\end{eqnarray*}
In view of Lemma~\ref{lem2.2}, the last expression can be
rewritten as
\begin{eqnarray*}
\lefteqn{a\sup_{\varphi\in S_{\kappa}}
       \left|\int_{\Omega}\bar\psi(y)(I_{m}\varphi)(y)\,dm(y)\right|^2} \\
&&\leq a \int_{\Omega}|\psi(y)|^2 dm(y)
      \sup_{\varphi\in S_{\kappa}}
      \underbrace{\int_{\Omega}|(I_{m}\varphi)(y)|\,dm(y)}_{\leq
      a\left<\varphi,\varphi\right>_{i\kappa}}
\leq a^2\int_{\Omega}|\psi(y)|^2 dm(y)
\end{eqnarray*}
where we have used Schwarz inequality and the assumption
(\ref{bass}) again. Consequently, $\|\alpha I_{m}
R_{m,dx}^{i\kappa}\|<a$ holds for all $\kappa\geq\kappa_{0}.\,$
\qed \vspace{3mm}

\noindent The resolvent expression of Proposition~\ref{resolv1}
represents a starting point for construction of the
Birman-Schwinger theory. This will be done in the next section for
our double-waveguide example. A part of the analysis is an
expression for the number of eigenvalues of $H_{\alpha m}$ which
can be derived in the present general context.
\begin{corol}\label{Cor.2.5}
$\dim\Ker(H_{\alpha m}-k^2)=\dim\Ker(I+\alpha I_{m}R_{m,dx}^{k})$
holds for any $k\in\Com_{\Omega,0}^+$.
\end{corol}
\PF
Suppose first that $\psi\in\sii(m)$ satisfies $\psi+\alpha
I_{m}R_{m,dx}^{k}\psi=0$. By Lemma~\ref{lem2.2} we have
\begin{eqnarray*}
&&{\cal E}_{\alpha m}(R_{m,dx}^{k}\psi,\varphi)
  -(k^2 R_{m,dx}^{k}\psi,\varphi)\\
&&\quad=\int_{\Omega}\bar\psi(y)(I_{m}\varphi)(y)\,dm(y)
  +\int_{\Omega}\alpha(y)(I_{m}\overline{R_{m,dx}^{k}\psi})(y)
                         (I_{m}\varphi)(y)\,dm(y)=0
\end{eqnarray*}
for all $\varphi\in\sobi(\Omega)\equiv D({\cal E}_{\alpha m})$.
By the second representation theorem, it follows that
$R_{m,dx}^{k}\psi\in D(H_{\alpha m})$ and
$H_{\alpha m}R_{m,dx}^{k}\psi=k^2 R_{m,dx}^{k}\psi$. Since $R_{m,dx}^{k}$
is injective by Lemma~\ref{lem2.2}, we get
\begin{displaymath}
  \dim\Ker(I+\alpha I_{m}R_{m,dx}^{k})\leq\dim\Ker(H_{\alpha m}-k^2).
\end{displaymath}
On the other hand, let $\varphi\in D(H_{\alpha m})$ with
$H_{\alpha m}\varphi=k^2\varphi$. In view of the above argument,
it is sufficient to show that $\varphi=R_{m,dx}^{k}\psi$ for some
$\psi\in\sii(m)$ such that $\psi+\alpha I_{m}R_{m,dx}^{k}\psi=0$.
We put $\psi:=-\alpha I_{m}\varphi$; then by Lemma~\ref{lem2.2},
we have
\begin{displaymath}
  {\cal E}_{0}(R_{m,dx}^{k}\psi,\chi)-(k^2 R_{m,dx}^{k}\psi,\chi)
  =-\int_{\Omega}\alpha(x)(I_{m}\bar\varphi)(x)(I_{m}\chi)(x)\,dm(x)
\end{displaymath}
for all $\chi\in\sobi(\Omega)$. Using the second representation
theorem again together with the assumption $H_{\alpha
m}\varphi=k^2\varphi$ we get ${\cal E}_{\alpha
m}(\varphi,\chi)-(k^2\varphi,\chi)=0$, so
\begin{eqnarray*}
\lefteqn{{\cal E}_{0}(\varphi,\chi)-(k^2\varphi,\chi)} \\
&&\qquad={\cal E}_{\alpha m}(\varphi,\chi)-(k^2\varphi,\chi)
  -\int_{\Omega}\alpha(x)(I_{m}\bar\varphi)(x)(I_{m}\chi)(x)\,dm(x) \\
&&\qquad=-\int_{\Omega}\alpha(x)(I_{m}\bar\varphi)(x)(I_{m}\chi)(x)\,dm(x)
\end{eqnarray*}
holds for any $\chi\in\sobi(\Omega)$. Comparing the two
expressions, we arrive at the relation $\varphi=R_{m,dx}^{k}\psi$
which yields $\psi+\alpha I_{m}R_{m,dx}^{k}\psi=\psi+\alpha
I_{m}\varphi=0.$ \qed

\setcounter{equation}{0}
\section{BS analysis for the double waveguide}
\label{Sec.BSex}
The core of the classical Birman-Schwinger analysis is a resolvent
expression containing at the l.h.s. only the free resolvent. If
the Schr\"odinger operator in question involves a potential
defined via a measure, the free resolvent has to be interpreted as
an operator between different $\sii$ spaces. Now we are going to
derive such a formula for the system described in
Sec.~\ref{model}.

\subsection{The basic lemma}
The sought relation is an analogue of \cite[Lemma~2.3]{BEKS} valid
for $\Omega=\Real^{d}$. The proof of this result employed the
explicit form of the resolvent, thus the argument had to be
modified again. We shall consider the operator $H_{\alpha m}$ of
the previous section in the situation when
$\Omega:=\Real\times\mathcal{O}$ and $m$ is supported by the
$x$-axis. Then we have:
\begin{lemma}\label{lem2.3}
 {\em (i)}   $\forall k\in\Com_{\Omega,0}^+:\ I_{m}R_{m,dx}^{k}=R_{m,m}^{k}$
and\ $I_{m}R_{0}^{k}=R_{dx,m}^{k}$. \\
 {\em (ii)}  $I+\alpha
R_{m,m}^{i\kappa}$ has a bounded inverse on $\sii(m)$ for all
$\kappa>0$ large enough. \\
 {\em (iii)} Assume that $I+\alpha
R_{m,m}^{k}$ is invertible for $k\in\Com_{\Omega,0}^+$ and the operator %
\begin{displaymath}
  R^{k}:=R_{0}^{k}-R_{m,dx}^{k}(I+\alpha R_{m,m}^{k})^{-1}
                   \alpha R_{dx,m}^{k}
\end{displaymath}
on $\sii(m)$ is everywhere defined. Then $k^2\in\rho(H_{\alpha
m})$ and $(H_{\alpha m}-k^2)^{-1}=R^{k}$. \\
 {\em (iv)}  $\forall k\in\Com_{\Omega,0}^+:\
\dim\Ker(H_{\alpha m}-k^2)=\dim\Ker(I+\alpha R_{m,m}^{k})$.
\end{lemma}
\PF Since the assertions (ii)--(iv) are easy consequences of the
first claim and the above corollaries, it is sufficient to check
(i). The free Green's function for the strip $\Omega$ was written
down in~\cite{EGST}. In particular, we have
\begin{eqnarray*}
  G_{0}(\vec x,\vec x';i\kappa)&=&\frac{1}{D}
  \sum\limits_{n=1}^{\infty}\frac{e^{-\kappa_{n}|x-x'|}}{\kappa_{n}}
  \sin\frac{\pi n}{D}(y+d_{2})\sin\frac{\pi n}{D}(y'+d_{2})\,;\\
 \kappa_{n}&:=&\sqrt{\kappa^2+\left(\frac{\pi n}{D}\right)^2}
\end{eqnarray*}
for $\kappa>0$, where $\vec x:=(x,y)$. We know that $G_{0}(\vec
x,\vec x';i\kappa)>0$, it is smooth in each argument,
exponentially decaying as $|x-x'|\to\infty$ and has a logarithmic
singularity as $\vec x'\to\vec x$. As in Ref.~\cite{BEKS}, we need
a smooth approximation to $G_{0}$. We employ the fact that
\begin{displaymath}
  G_{0}(\vec x, \vec x';i\kappa)
  =-\frac{1}{2\pi}\ln|\vec x-\vec x'|+\Gamma(\vec x,\vec x')\,,
\end{displaymath}
where $\Gamma$ is a $\Smooth$ function vanishing when $y,y'$
assume the values $d_{1},-d_{2}$. We take
\begin{itemize}
\item a strictly increasing $\Smooth$ function $\xi:(0,\infty)
\to[1,\infty)$ such that $\xi(0)=1$ and $\xi(x)=x$ for $x\geq 2$,
\vspace{-1.2ex}
\item an increasing sequence $\{\eta_{n}\}_{n=1}^{\infty}
\subset\Comp(\Omega)$ such that $\lim_{n\to\infty}\eta_{n}(\vec
x)=1$ for any fixed $\vec x\in\Omega$ \vspace{-1.2ex}
\end{itemize}
and put
\begin{displaymath}
  G_{n}(\vec x, \vec x';i\kappa)
  :=\left[-\frac{1}{2\pi}\ln\frac{\xi(n|\vec x-\vec x'|)}{n}
  +\Gamma(\vec x,\vec x')\right]\eta_{n}(\vec x).
\end{displaymath}
Clearly,
\begin{eqnarray*}
\mbox{(i)}  && G_{n}(\cdot,\vec x;i\kappa)\in\Comp(\Omega) \\
\mbox{(ii)} && \forall\, \vec x,\vec x'\in\Omega:\; \quad
G_{n}(\vec x,\vec x';i\kappa) \leq G_{n+1}(\vec x,\vec x';i\kappa)
\\
\mbox{(iii)}&& \exists\, c_{1}>0\;\: \forall\,\vec x,\vec
x'\in\Omega,\:\vec x\not=\vec x':\quad |\nabla_{x}G_{n}(\vec
x,\vec x';i\kappa)| \leq\frac{c_{1}}{|\vec x-\vec x'|} \\
\mbox{(iv)} && \exists\, c_{2},c_{3}>0\;\: \forall\,\vec x,\vec
x'\in\Omega, \:|\vec x-\vec x'|\; \mbox{\rm large enough}: \\
&&\qquad\qquad |G_{n}(\vec x,\vec x';i\kappa)|
+|\nabla_{x}G_{n}(\vec x,\vec x';i\kappa)| \leq
c_{2}e^{-c_{3}|x-x'|}
\end{eqnarray*}
We use the common notation $\mu$ for $m,dx$. Take an arbitrary
$\psi\in\sii(\mu)$ and $\varphi_{n}:=R_{\mu,dx}^{n,i\kappa}\psi$,
\ie
\begin{displaymath}
  \varphi_{n}=\int_{\Omega}G_{n}(\vec x,\vec x';i\kappa)
  \psi(\vec x)\,d\mu(\vec x').
\end{displaymath}
Each $\varphi_{n}\in\sobi(\Omega)$ by definition.
Furthermore, by the construction of the regularized
Green's function, we have also $\varphi_{n}\in\Comp(\Omega)$.
Next we have to estimate the Sobolev norm of $\varphi_{n}$.
In view of (iii) we have
\begin{eqnarray*}
\lefteqn{\int_{\Omega}\left|\nabla_{x}\int_{\Omega}G_{n}
  (\vec x,\vec x';i\kappa)\psi(x')\,d\mu(x')\right|^2 dx} \\
&&=\int_{\Omega}
  \left|\int_{\Omega}\nabla_{x}G_{n}(\vec x,\vec x';i\kappa)
  \psi(x')\,d\mu(x')\right|^2 dx \\
&&\leq\int_{\Omega}\left|\int_{\Omega}\frac{c_{1}}{|\vec x-\vec x'|}
  \psi(x')\,d\mu(x')\right|^2 dx
\leq\int_{\Real^2}\left|\int_{\Real^2}\frac{c_{1}}{|\vec x-\vec x'|}
  \psi(x')\,d\mu(x')\right|^2 dx
\end{eqnarray*}
where $\psi$ in the last expression means the trivial extension
from $\Omega$ to $\Real^2$. The integral was shown to be finite in
the proof of Lemma~2.3 in~\cite{BEKS}. As for the non-derivative
part we notice that $G_{0}$ has a bound as a consequence of the
fact that $R_{\mu,dx}^{i\kappa}\psi\in
\sobi(\Omega)\subset\sii(\Omega)$.

Summing the above considerations, we have demonstrated that
$\{\varphi_{n}\}$ is a bounded sequence in the local Sobolev space
$\sobi(\Omega)$. Then we proceed as above: we construct
$\chi_{n}:=\frac{1}{n}\sum_{j=1}^{n}\varphi_{j}$ and use the
diagonal trick to show that (a subsequence of) $\{\chi_{n}\}$
converges strongly in $\sobi(\Omega)$. By the property (ii) and
monotone convergence theorem
\begin{displaymath}
  \forall\vec x\in\Omega:\quad
  \lim_{n\to\infty}\chi_{n}(\vec x)=(R_{\mu,dx}^{i\kappa}\psi)(\vec x),
\end{displaymath}
so $\chi_{n}\to R_{\mu,dx}^{i\kappa}\psi$ in $\sobi(\Omega)$ as
$n\to\infty$. From the definition of $I_{m}$,
\begin{displaymath}
  I_{m}\chi_{n}\to I_{m}R_{\mu,dx}^{i\kappa}\psi\quad\mbox{in}\;\, \sii(m)
\end{displaymath}
as $n\to\infty$. Since $I_{m}\chi_{n}\in\Comp(\Omega)$ by
construction, we conclude that $I_{m}\chi_{n}=\chi_{n}$ holds
$m\!-\!\alev$, and therefore
\begin{displaymath}
  I_{m}R_{\mu,dx}^{i\kappa}\psi=R_{\mu,dx}^{i\kappa}\psi
  \quad m\!-\!\alev
\end{displaymath}
This proves the desired relations for $k$ purely imaginary. The
result extends to any $k\in\Com_{\Omega,0}^+$ by means of the
Hilbert identity as in the proof of Lemma~\ref{lem2.2}.\qed

\subsection{The resolvent comparison formula}

From the point of view of our model the formula in
Lemma~\ref{lem2.3}(iii) still suffers from two defects. First of
all, if we consider a semitransparent barrier whose coupling
parameter is varied locally, it is natural to take $H_{\alpha m}$
with a constant but generally nonzero $\alpha$ as a comparison
operator. Secondly, in analogy with the classical Birman-Schwinger
theory it is useful to arrange the ``potential" symmetrically with
respect to the free resolvent in order to be able to use
efficiently its decay properties.

Let $\alpha$ be as above a Borel measurable function
$\Real\to\Real$ and $\alpha_{0}\in\Real$; abusing the notation we
shall employ the symbol $\alpha_{0}$ also for the constant
function $\Real\to\Real$, $\alpha_{0}(x)=\alpha_{0}$. By the
preceding result, we have
$$
 R^{k}(\alpha_{0})
 =R_{0}^{k}-R_{m,dx}^{k}(I+\alpha_{0}R_{m,m}^{k})^{-1}
 \alpha_{0}R_{dx,m}^{k},
$$
so for any~$k\in\Com_{\Omega,0}^+$ with $k^2\in\rho(H_{\alpha_0 m})
\cap\rho(H_{\alpha m})$:
\begin{eqnarray}\label{resdif}
R^{k}(\alpha)\!-\!R^{k}(\alpha_{0}) &\!=\!&
  R_{m,dx}^{k}\left[(I+\alpha_{0}R_{m,m}^{k})^{-1}\alpha_{0}-
  (I+\alpha R_{m,m}^{k})^{-1}\alpha\right]R_{dx,m}^{k}\nonumber\\
&\!=\!& R_{m,dx}^{k}(I+\alpha R_{m,m}^{k})^{-1}
  \left[\alpha_{0}(I+\alpha R_{m,m}^{k})-
  \alpha(I+\alpha_{0}R_{m,m}^{k})\right] \nonumber \\
&& \times\,(I+\alpha_{0}R_{m,m}^{k})^{-1}R_{dx,m}^{k}\nonumber\\
&\!=\!& R_{m,dx}^{k}(I+\alpha
R_{m,m}^{k})^{-1}(\alpha_{0}\!-\!\alpha)
  (I+\alpha_{0}R_{m,m}^{k})^{-1}R_{dx,m}^{k}\,,
\end{eqnarray}
where in the second line we have used the fact that $\alpha_{0}$
is a number and thus commutes with $I+\alpha R_{m,m}^{k}$. Next we
compute traces of $R^{k}(\alpha_{0})$. By Lemma~\ref{lem2.3}, we
have
\begin{eqnarray*}
  R_{dx,m}^{k}(\alpha_{0})
&=&R_{dx,m}^{k}
  -R_{m,m}^{k}(I+\alpha_{0}R_{m,m}^{k})^{-1}\alpha_{0}R_{dx,m}^{k}\\
&=&(I+\alpha_{0}R_{m,m}^{k})^{-1}R_{dx,m}^{k}.
\end{eqnarray*}
On the other hand, $(R_{\mu,\nu}^{k})^{*}$ maps $\sii(\nu)$ into
$\sii(\mu)$ and $(R_{\mu,\nu}^{k})^{*}=R_{\nu,\mu}^{\bar k}$. In
the same way as above, this yields
$$
  R_{m,dx}^{k}(\alpha_{0}) = R_{m,dx}^{k}(I+\alpha_{0}
  R_{m,m}^{k})^{-1}\,;
$$
applying once more Lemma~\ref{lem2.3} we get also
\begin{equation}\label{NEW}
  R_{m,m}^{k}(\alpha_{0})=(I+\alpha_{0}R_{m,m}^{k})^{-1}R_{m,m}^{k}
  =R_{m,m}^{k}(I+\alpha_{0}R_{m,m}^{k})^{-1}.
\end{equation}
We employ these relations to proceed with the calculation of the
resolvent difference~(\ref{resdif}): up to a sign change it equals
\begin{displaymath}
  R_{m,dx}^k(\alpha_0)(I+\alpha_0 R_{m,m}^k)(I+\alpha R_{m,m}^k)^{-1}
  (\alpha\!-\!\alpha_0)R_{dx,m}^k(\alpha_0)
\end{displaymath}
and the central expression can be further rewritten as follows:
\begin{eqnarray}\label{thirdline}
\lefteqn{(I+\alpha_0 R_{m,m}^k)(I+\alpha R_{m,m}^k)^{-1}
  (\alpha\!-\!\alpha_0)}\nonumber\\
&&=\left[\left(I+\alpha_0 R_{m,m}^k
  +(\alpha\!-\!\alpha_0)R_{m,m}^k\right)
  (I+\alpha_0 R_{m,m}^k)^{-1}\right]^{-1}(\alpha\!-\!\alpha_0)\nonumber\\
&&=\left[I+(\alpha\!-\!\alpha_0)R_{m,m}^k(\alpha_0)\right]^{-1}
  (\alpha\!-\!\alpha_0)\nonumber\\
&&=\left[I+(\alpha\!-\!\alpha_0)R_{m,m}^k(\alpha_0)\right]^{-1}
  (\alpha\!-\!\alpha_0)^\frac{1}{2}
  \left[I+|\alpha\!-\!\alpha_0|^{\frac{1}{2}}R_{m,m}^k(\alpha_0)
  (\alpha\!-\!\alpha_0)^{\frac{1}{2}}\right]\nonumber\\
&&\quad\times
  \left[I+|\alpha\!-\!\alpha_0|^{\frac{1}{2}}R_{m,m}^k(\alpha_0)
  (\alpha\!-\!\alpha_0)^{\frac{1}{2}}\right]^{-1}
   |\alpha\!-\!\alpha_0|^\frac{1}{2}\nonumber\\
&&=\left[I+(\alpha\!-\!\alpha_0)R_{m,m}^k(\alpha_0)\right]^{-1}
  \left[I+(\alpha\!-\!\alpha_0)R_{m,m}^k(\alpha_0)\right]
   (\alpha\!-\!\alpha_0)^\frac{1}{2}\nonumber\\
&&\quad\times
  \left[I+|\alpha\!-\!\alpha_0|^{\frac{1}{2}}R_{m,m}^k(\alpha_0)
  (\alpha\!-\!\alpha_0)^{\frac{1}{2}}\right]^{-1}
  |\alpha\!-\!\alpha_0|^{\frac{1}{2}}\nonumber\\
&&=(\alpha\!-\!\alpha_0)^{\frac{1}{2}}
  \left[I+|\alpha\!-\!\alpha_0|^{\frac{1}{2}}R_{m,m}^k(\alpha_0)
  (\alpha\!-\!\alpha_0)^{\frac{1}{2}}\right]^{-1}
  |\alpha\!-\!\alpha_0|^{\frac{1}{2}}\,.
\end{eqnarray}
We employ here the usual ``square-root convention" of the
Birman-Schwinger theory, $(\alpha\!-\!\alpha_0)^{\frac{1}{2}}
:=|\alpha\!-\!\alpha_0|^{\frac{1}{2}} \sgn(\alpha\!-\!\alpha_0)$.

Notice finally that the obtained expression no longer contains
$R^k_0$. Using once again the Hilbert-identity trick, we can
extend its validity to the set $\Com_{\Omega,\alpha_0}^+ := \left\{k:
\: \im k>0\, \;\mathrm{or}\;k^2\in \left[0,\inf\sigma(H_{\alpha_0
m})\right)\right\}$. Summing up the above discussion, we get
\begin{thm}\label{thm3.1}
Under the stated assumptions, the resolvent of $H_{\alpha m}$ can
be expressed by means of that of the reference operator
$H_{\alpha_{0} m}$ as
\begin{eqnarray}\label{resolvent}
  R^{k}(\alpha) &=& R^{k}(\alpha_{0})
  -R_{m,dx}^{k}(\alpha_0)\,(\alpha\!-\!\alpha_0)^{\frac{1}{2}}\nonumber\\
&&\times\left[I+|\alpha\!-\!\alpha_0|^{\frac{1}{2}}\,R_{m,m}^k(\alpha_0)\,
  (\alpha\!-\!\alpha_0)^{\frac{1}{2}}\right]^{-1}
  |\alpha\!-\!\alpha_0|^{\frac{1}{2}}
  R_{dx,m}^{k}(\alpha_0)\nonumber
\end{eqnarray}
for any $k\in\Com_{\Omega,\alpha_0}^+$.
\end{thm}
Hence the original problem is equivalent to spectral analysis of
the integral operator
\beq\label{opK}
K_{\alpha}^k:=|\alpha\!-\!\alpha_0|^{\frac{1}{2}}\,R_{m,m}^k(\alpha_0)\,
(\alpha\!-\!\alpha_0)^{\frac{1}{2}}\,. \eeq
To be more specific we restrict ourselves to the situation which
we shall discuss below and adopt the following assumptions:
\begin{description}
  \item[\mbox{\sf (a1)}]
       $\:\alpha(\cdot)\!-\!\alpha_{0}\in\sieps(\Real,dx)$\quad
       for some $\varepsilon>0$,
  \item[\mbox{\sf (a2)}]
       $\:\alpha(\cdot)\!-\!\alpha_{0}\in\si(\Real,|x|\,dx)$;
\end{description}
notice that as a consequence of \mbox{\sf (a1)}, \mbox{\sf (a2)},
the function belongs also to $\si(\Real,dx)$.
\begin{corol}\label{corol3.4}
Under the assumption {\sf (a1), (a2)}, \\ {\em (i)} $K_\alpha^k$
is Hilbert-Schmidt for any $k\in\Com_{\Omega,\alpha_0}^+$, \\ {\em
(ii)} $I+K_\alpha^{i\kappa}$ has a bounded inverse on $\sii(m)$
for all $\kappa>0$ large enough, \\ {\em (iii)} $\forall
k\in\Com_{\Omega,\alpha_0}^+:\ \dim\Ker(H_{\alpha
m}-k^2)=\dim\Ker\left(I+K_\alpha^k\right),$ \\ {\em (iv)}
Birman-Schwinger principle holds, \label{K=-1}
$$
  \forall k\in\Com_{\Omega,\alpha_0}^+:\
  k^2\in\sigma_\mathrm{disc}(H_{\alpha})
  \Longleftrightarrow
  -1\in\sigma_\mathrm{disc}(K_{\alpha}^{k})\,.
$$
\end{corol}
\PF The Hilbert-Schmidt property follows by an argument analogous
to the estimate of $N_0^1$ in Proposition~\ref{HS-lims} below,
with the difference that the summation includes the first
transverse mode too and $\sqrt{\nu_n-\nu_1}$ is replaced by
$\sqrt{\nu_n-k^2}$. To prove (ii) one employs the analogue of
(\ref{est-sum}) to infer that $\|K_\alpha^{i\kappa}\|^2_{HS} \to
0$ as $\kappa\to\infty$.

To deal with the rest, notice first that it is sufficient to prove
(iii), (iv) for functions $\alpha$ which are essentially bounded.
Indeed, define $\alpha_N(x):= \mathrm{sgn\,}\alpha(x)\, \min \{
|\alpha(x)|,N \}$. By absolute continuity of the Lebesgue integral
the values of the quadratic form (\ref{formE}) related to
$\alpha_N$ converge to that of ${\cal E}_{\alpha m}$ as $N\to
\infty$ so $H_{\alpha_N m}\to H_{\alpha m}$ in the strong
resolvent sense by \cite[Thm. VIII.3.6]{Ka}. Hence the discrete
spectrum of $H_{\alpha m}$ is approximated by that of $H_{\alpha_N
m}$ \cite[Sec.~VIII.7]{RS} but the latter has a finite dimension
bound uniformly w.r.t. $N$ as we shall show in
Proposition~\ref{num-est}.

Suppose therefore that $\|\alpha\!-\!\alpha_0\|_{\infty} <
\infty$. The operator $I+K_\alpha^k$ has by~(i) a purely discrete
spectrum, every non-unit eigenvalue being of a finite
multiplicity. Consequently, if $K_\alpha^k$ has the eigenvalue
$-1$, the number $k^2$ belongs to the spectrum of $H_{\alpha m}$
with the same multiplicity. On the other hand, if there is no
$\psi$ solving $K_{\alpha}^k\psi=-\psi$, then
$(I+K_\alpha^k)^{-1}$ is bounded, and so is $R^k(\alpha)$, thus
$k^2\in \rho(H_{\alpha m})$. \qed

\begin{rems}{\rm
{\em (i)} It is clear from~(\ref{thirdline}) that there are other
expressions of the resolvent, \eg,
\begin{equation}\label{nonsymresolvent}
  R^{k}(\alpha)=R^{k}(\alpha_{0})-R_{m,dx}^{k}(\alpha_0)
  \left[I+(\alpha\!-\!\alpha_0)R_{m,m}^k(\alpha_0)\right]^{-1}
  (\alpha\!-\!\alpha_0)R_{dx,m}^{k}(\alpha_0)\,.
\end{equation}
The advantage of the fully symmetric form is that it allows an
optimal use of the decay properties of $\alpha\!-\!\alpha_0$. For
instance, in the weak-coupling analysis of the next section the
relation~(\ref{nonsymresolvent}) would force us to restrict
ourselves to the compact-support case. \\
{\em (ii)} As usual the BS principle provides an information about
the discrete spectrum. Whether eigenvalues embedded in the
essential spectrum may exist in the present situation remains an
interesting open problem.}
\end{rems}

\setcounter{equation}{0}
\section{Weak coupling}\label{Sec.WeakCoupling}
Now we shall apply the above general results to weak-coupling
analysis of our model represented by the Hamiltonian~$H_\alpha$
defined in~Section~\ref{Sec.BS1} (from now on we will omit the
subscript~$m$) which is considered as a perturbation of
an~$H_{\alpha_0}$ with a constant~$\alpha_0$. The spectral
properties of the latter operator are found easily; the
corresponding analysis was done in~\cite[Sec.~2.2]{EK1} where the
reader can find a detailed account.

We have seen that the function
$\alpha\!-\!\alpha_0$ plays the role of an effective potential. To
introduce a parameter controlling the perturbation, we replace it
in this section by $\lambda(\alpha\!-\!\alpha_0)$ with a small
$\lambda$. Without loss of generality, we may suppose that the
parameter is positive.

We shall concentrate on the case when the ``leaky part" is
localized in the sense that $\alpha(x)\!-\!\alpha_0$ decays fast
enough as $|x|\to\infty$. Using a variational argument we have
shown in \cite{EK1} that the discrete spectrum is then non-empty.
For a small positive $\lambda$ there is a unique bound state; our
aim here is to derive an asymptotic expansion of the corresponding
eigenvalue. The method follows the standard argument for
one-dimensional Schr\"odinger operators \cite{BGS, Si} and its
extension to waveguide systems \cite[Thm.~4.2.]{DE1}. We shall
suppose in the following that the assumptions {\sf (a1)} and
{\sf (a2)} are valid.

\subsection{Preliminaries}
In the following we denote the ground-state eigenvalue of the
weakly coupled Hamiltonian as $k^2$ and look for the function
$\lambda\mapsto k^2$. Since we are interested in the discrete
spectrum, we consider the Green's function for
$k^2\!<\!\nu_{1}(\alpha_{0})$, the threshold of the essential
spectrum given by the first transverse eigenvalue of the
unperturbed system -- \cf~\cite[Sec.~3]{EK1}.
\begin{equation}\label{kernel}
  K_{\alpha}(x,x';k)=|\alpha(x)\!-\!\alpha_{0}|^{\frac{1}{2}}
  \sum_{n=1}^{\infty}\frac{|\chi_{n}(0;\alpha_{0})|^2 }{2\kappa_{n}}
  \,e^{-\kappa_{n}|x-x'|}
  \ (\alpha(x')\!-\!\alpha_{0})^{\frac{1}{2}}\,,
\end{equation}
where $\kappa_{n}:=\sqrt{\nu_{n}(\alpha_{0})-k^2}$ and
$\{\chi_n\}$ is the family the corresponding transverse
eigenfunctions of the unperturbed system. It is straightforward to
check that $\lim_{\lambda\to 0}k^2=\nu_{1}$, \ie, $\kappa_{1}\!\to
0$ as $\lambda\to 0$.

The key idea of the following argument is that $K_{\alpha}^k$ is
well behaved in the limit $k^2\to\nu_1$ except for a divergent
rank-one part. The singularity is contained in the first term of
the expansion~(\ref{kernel}) and can be singled out by taking (as
in) $K_{\alpha}^k=Q_{\alpha}+P_{\alpha} =Q_{\alpha}
+A_{\alpha}+N_{\alpha}$ in analogy with ~\cite{BGS, DE1}, where
\begin{eqnarray}\label{QAN}
  Q_{\alpha}(x,x') &=& |\alpha(x)\!-\!\alpha_{0}|^{\frac{1}{2}}
                       \,e^{-\kappa_1 |x|}
                       \ \frac{|\chi_{1}(0)|^2}{2\kappa_{1}}
                       \,e^{-\kappa_1 |x'|}
                       \ (\alpha(x')\!-\!\alpha_{0})^{\frac{1}{2}}
                       \nonumber\\
  A_{\alpha}(x,x') &=& |\alpha(x)\!-\!\alpha_{0}|^{\frac{1}{2}}
                       \ \frac{|\chi_{1}(0)|^2}{\kappa_{1}}
                       \,e^{-\kappa_1 |x|_>}
                       \sinh\kappa_1 |x|_<
                       \ (\alpha(x')\!-\!\alpha_{0})^{\frac{1}{2}}
                       \qquad\\
  N_{\alpha}(x,x') &=& |\alpha(x)\!-\!\alpha_{0}|^{\frac{1}{2}}
                       \sum_{n=2}^{\infty}
                       \frac{|\chi_{n}(0)|^2}{2\kappa_{n}}
                       \,e^{-\kappa_{n}|x-x'|}
                       \,(\alpha(x')\!-\!\alpha_{0})^{\frac{1}{2}}\,.
                       \nonumber
\end{eqnarray}
We have introduced here $|x|_<:=\max\left\{0,
\min\{|x|,|x'|\}\sgn(xx')\right\}$ and $|x|_>:=\max\{|x|,|x'|\}$;
for the sake of brevity we drop $\alpha_0$ from the argument of
$\chi_n$. Defining
\begin{eqnarray*}
  A_{0}(x,x') &:=& |\alpha(x)\!-\!\alpha_{0}|^{\frac{1}{2}}
    \ |\chi_{1}(0)|^2\ |x|_<
    \ (\alpha(x')\!-\!\alpha_{0})^{\frac{1}{2}}\\
  N_{0}^{\beta}(x,x') &:=& |\alpha(x)\!-\!\alpha_{0}|^{\frac{1}{2}}
    \sum_{n=2}^{\infty}\frac{|\chi_{n}(0)|^2}{2}\
    \frac{e^{-\beta\sqrt{\nu_{n}-\nu_{1}}|x-x'|}}{\beta\sqrt{\nu_{n}-\nu_{1}}}
    \ (\alpha(x')\!-\!\alpha_{0})^{\frac{1}{2}}
\end{eqnarray*}
with $\beta>0$, we get
\begin{prop}\label{HS-lims}
Let the assumptions \mbox{\sf (a1)}, \mbox{\sf (a2)} be valid,
then
\begin{equation}
  \lim_{\kappa_{1}\to 0}\|A_{\alpha}-A_{0}\|_{HS}=0
  \qquad\mbox{\rm and}\qquad
  \lim_{\kappa_{1}\to 0}\|N_{\alpha}-N_{0}^1\|_{HS}=0\,.
\end{equation}
\end{prop}
\PF $\;A_{0}$ is Hilbert-Schmidt since
\begin{eqnarray*}
  \|A_{0}\|_{HS}^2 & =  & \int_{\Real^2}|A_{0}(x,x')|^2\,dx\,dx'\\
                   & =  & |\chi_{1}(0)|^4\int_{\Real^2}
                          |\alpha(x)\!-\!\alpha_{0}|\ |x|_<^2\
                          |\alpha(x')\!-\!\alpha_{0}|\,dx\,dx'\\
                   &\leq& |\chi_{1}(0)|^4\left(\int_{\Real}
                          |x|\ |\alpha(x)\!-\!\alpha_{0}|\,dx\right)^2
                          <\infty
\end{eqnarray*}
by assumption. We have $\lim_{\kappa_{1}\to 0}A_{\alpha}=A_{0}$
and $|A_{\alpha}(x,x';\kappa_{1})|\leq|A_{0}(x,x')|$. This allows
us to use the dominated convergence theorem which yields
immediately the first claim.

$N_0^1$ has a logarithmic singularity as $x'\to x$. Nevertheless,
its Hilbert-Schmidt norm is finite because
\begin{eqnarray*}
  \|N_{0}^1\|_{HS}^2
&=&
  \sum_{m,n=2}^{\infty}
  \frac{|\chi_{n}(0)\,\chi_{m}(0)|^2}
  {4\sqrt{\nu_{n}-\nu_{1}}\sqrt{\nu_{m}-\nu_{1}}}\\
&&
  \times \int_{\Real^2}\ |\alpha(x)\!-\!\alpha_{0}|\
  e^{-(\sqrt{\nu_{n}-\nu_{1}}+\sqrt{\nu_{m}-\nu_{1}})|x-x'|} \
 |\alpha(x')\!-\!\alpha_{0}|\ dx\,dx' \,,
\end{eqnarray*}
where the monotone convergence theorem justifies the interchange
of summation and integration, and by H\"older inequality the
integral can be estimated by
\begin{eqnarray*}
\lefteqn{\|\alpha\!-\!\alpha_{0}\|_{1+\varepsilon}
  \int_{\Real} dx \, |\alpha(x)\!-\!\alpha_{0}| } \\
&&
  \times
  \Bigg\{ e^{-(\sqrt{\nu_{n}-\nu_{1}}+\sqrt{\nu_{m}-\nu_{1}}) x}
  \left( \int_{-\infty}^x
  e^{(\varepsilon')^{-1}
  (\sqrt{\nu_{n}-\nu_{1}}+\sqrt{\nu_{m}-\nu_{1}}) x'}
  dx'\right)^{\varepsilon'} \\
&&
  \phantom{\times}
  +e^{(\sqrt{\nu_{n}-\nu_{1}}+\sqrt{\nu_{m}-\nu_{1}}) x}
  \left( \int_x^\infty
  e^{-(\varepsilon')^{-1}
  (\sqrt{\nu_{n}-\nu_{1}}+\sqrt{\nu_{m}-\nu_{1}}) x'}
  dx' \right)^{\varepsilon'}
  \Bigg\} \\
&=&
  2 (\varepsilon')^{\varepsilon'}
  \|\alpha\!-\!\alpha_{0}\|_{1+\varepsilon} \|\alpha\!-\!\alpha_{0}\|_1
  (\sqrt{\nu_{n}-\nu_{1}}+\sqrt{\nu_{m}-\nu_{1}})^{-\varepsilon'},
\end{eqnarray*}
where $\varepsilon'$ is an abbreviation for
$\frac{\varepsilon}{1+\varepsilon}>0$. The
sequence~$\{\chi_n(0)\}$ is uniformly bounded. To see this one has
to employ the explicit expression of the transverse eigenfunctions
given by the relations (2.7) and (2.8) of \cite{EK1}. It implies
the estimate $\chi_n(0)^2\leq 2 h_1(u) h_2(u)$, where
$u:=\sqrt{\nu_n}$ and
$$
  h_j(u):=\frac{\sqrt{u}\,|\sin{d_j u}|}
  {\sqrt{2 d_j u-\sin 2 d_j u}}
$$
for $j=1,2$. These functions are bounded in $(0,\infty)$ because
they are continuous inside the interval and the limits
$$
  \lim_{u\to 0+} h_j(u)=(2d_j)^{-\frac{1}{2}}
  \qquad
  \limsup_{u\to\infty} h_j(u)=\left(3/(2d_j)\right)^{-\frac{1}{2}}
$$
are finite. Consequently, it is sufficient to check the
convergence of
\begin{equation} \label{est-sum}
  \sum_{m,n=2}^{\infty}
  \frac{1}{\sqrt{\nu_{n}-\nu_{1}}\sqrt{\nu_{m}-\nu_{1}}
  \,(\sqrt{\nu_{n}-\nu_{1}}
  +\sqrt{\nu_{m}-\nu_{1}})^{\varepsilon'}}\,,
\end{equation}
however, $\nu_n^{-1/2}=o(n^{-1})$ as $n\to\infty$ --
\cf~\cite[Lemma~2.2]{EK1}. The assumptions of the dominated
convergence theorem are fulfilled again because
$\lim_{\kappa_{1}\to 0}N_{\alpha}=N_{0}^1$ and
$|N_{\alpha}(x,x';\kappa_{1})|\leq|N_{0}^1(x,x')|$.\qed
\medskip

We will also need the boundedness of $A_{\alpha}$ and $N_{\alpha}$
for complex $z:=\kappa_1$. It follows from the above results,
since it is easy to prove
\begin{lemma}\label{boundAN}
  {\rm (i)} $\quad \forall\, z\!\in\!\Com,\, \re z\geq 0,\,
  |z|\!<\!\frac{\pi}{2}:\ |A_{\alpha}(z)|\leq\sqrt{2}\,|A_{0}|$
  \smallskip

  {\rm (ii)} $\quad \exists\, C,\beta>0\quad
  \forall\, z\!\in\!\Com,\, \re z\geq 0,|z|\!<\!\sqrt{\nu_{n}-\nu_{1}}:\
  |N_{\alpha}(z)|\leq C |N_{0}^{\beta}|\,$.
\end{lemma}

\subsection{Existence of the ground state}
Now we want to prove the following basic result:
\begin{thm}\label{mainThm}
Assume that the hypotheses \mbox{\sf (a1)}, \mbox{\sf (a2)} are
valid. Then $H_{\alpha}$ has at most one simple eigenvalue
$E(\lambda)<\nu_{1}$ for small enough $\lambda$, and this happens
if and only if the equation
\begin{equation}\label{mainthm}
  \kappa_{1}=
  -\frac{\lambda}{2}\,|\chi_{1}(0)|^2
  \left(\,e^{-\kappa_1|\cdot|}\,(\alpha\!-\!\alpha_0)^{\frac{1}{2}},
  (I+\lambda P_{\alpha})^{-1}\,e^{-\kappa_1|\cdot|}\,
  |\alpha\!-\!\alpha_{0}|^{\frac{1}{2}}\,\right)
\end{equation}
for $\kappa_{1}:= \sqrt{\nu_{1}-E}$ has a positive solution.
\end{thm}
\PF It is clear from the proof of Proposition~\ref{HS-lims} that
$\|P_{\alpha}\|\leq\|A_{0}\|_{HS}+\|N_{0}\|_{HS}<\infty$, thus
$\|\lambda P_{\alpha}\|<1$ holds for sufficiently small $\lambda$.
Then $I+\lambda P_{\alpha}$ is invertible and we may write
\begin{displaymath}
  (I+\lambda K_{\alpha}^k)^{-1}=
  \left[I+(I+\lambda P_{\alpha})^{-1}\lambda Q_{\alpha}\right]^{-1}
  (I+\lambda P_{\alpha})^{-1}.
\end{displaymath}
It follows that $\lambda K_{\alpha}$ has eigenvalue $-1$ if and
only if the same is true for $(I+\lambda P_{\alpha})^{-1}\lambda
Q_{\alpha}$. Since $Q_{\alpha}$ is a rank-one operator by
(\ref{QAN}), we can express it as $(I+\lambda
P_{\alpha})^{-1}\lambda Q_{\alpha}=(\psi,\cdot)\,\varphi$ with
$\psi:=\lambda\ \frac{|\chi_1(0)|^2}{2\kappa_1}\ e^{-\kappa_1
|\cdot|}\ (\alpha(\cdot)-\alpha_{0})^{\frac{1}{2}}$ and
$\varphi:=(I+\lambda P_{\alpha})^{-1}\ e^{-\kappa_1 |\cdot|}\
|\alpha(\cdot)-\alpha_{0}|^{\frac{1}{2}}$; it has just one
eigenvalue, namely $(\psi,\varphi)$. Putting it equal to $-1$ we
get the condition (\ref{mainthm}).

This proves the theorem except for the assertion that
(\ref{mainthm}) has {\em at most} one positive solution for
$\lambda$ small and fixed. Since there is a one-to-one
correspondence between eigenvalues of $H_{\alpha}$ and solutions
of (\ref{mainthm}) and the number of eigenvalue cannot decrease
after the replacement $\alpha\!-\!\alpha_{0}\mapsto
-|\alpha\!-\!\alpha_{0}|$, we need only show that (\ref{mainthm})
has at most one solution when $\alpha\!-\!\alpha_{0}\leq 0$. In
this case, (\ref{mainthm}) is equivalent to
\mbox{$z=G(z,\lambda)$} where
\begin{equation}\label{G}
  G(z,\lambda):=\frac{\lambda}{2}\,|\chi_{1}(0)|^2
  \left(\,e^{-z|\cdot|}\,|\alpha\!-\!\alpha_0|^{\frac{1}{2}},
  (I+\lambda P_{\alpha})^{-1}\,e^{-z|\cdot|}\,
  |\alpha\!-\!\alpha_{0}|^{\frac{1}{2}}\,\right).
\end{equation}

To complete the proof, we need several lemmas. The symbols $C,
C_j$ in the following are unspecified constants.
\begin{lemma}\label{AuxLem1}
If $\alpha\!-\!\alpha_{0}\not\equiv 0$, then $|z|^{-1}\leq
C_{1}\lambda^{-1}$ holds for $\lambda$ small.
\end{lemma}
\PF From~(\ref{G}) we see that any solution of
\mbox{$z=G(z,\lambda)$} for $\lambda$ small must obey
\mbox{$z=\frac{\lambda}{2}|\chi_{1}(0)|^2
\int_{\Real}|\alpha(x)\!-\!\alpha_{0}|\,dx+{\cal O}(\lambda^2)$},
which yields the assertion provided $\alpha\!-\!\alpha_{0}$ is not
identically zero. \qed

\begin{lemma}\label{AuxLem2}
For sufficiently small $z$, \mbox{ $\left\|\frac{\partial
P_{\alpha}}{\partial z}\right\| < C_{2}|z|^{-1}$}.
\end{lemma}
\PF Let us choose a circular contour, $\varphi:s=z(1+e^{it}),\;
t\in[0,2\pi)$. The operator-valued fuction $P_{\alpha}(\cdot)$ is
real-analytic in the region $\re z>0$ and has a bounded limit as
$z\to 0+$. Hence Cauchy integral formula together with
Lemma~\ref{boundAN} gives
\begin{displaymath}
  \left|\frac{\partial P_{\alpha}}{\partial z}\right|=
  \left|\frac{1}{2\pi i}\int_{0}^{2\pi}
  \frac{P_{\alpha}(s)\,iz~e^{i t} dt}{z^2\,e^{2 i t}}\right|
  \leq\frac{\sqrt{2}|A_{0}|+C|N_{0}^{\beta}|}{|z|}\,,
\end{displaymath}
but $A_{0},N_{0}^{\beta}$ have finite HS norms.\qed

\begin{lemma}\label{AuxLem3}
For any $z_{0}\in\Real$ there is $C_{3}>0$ such that for all
$z\in[0,z_{0}]$ we have
\begin{displaymath}
  \quad\left|\left(
  \,e^{-z|\cdot|}\,|\alpha\!-\!\alpha_0|^{\frac{1}{2}},
  \left[2\,|\cdot|+\frac{\partial P_{\alpha}}{\partial z}\right]
  \,e^{-z|\cdot|}\,|\alpha\!-\!\alpha_{0}|^{\frac{1}{2}}\,
  \right)\right|\leq C_{3}
\end{displaymath}
\end{lemma}
\PF An explicit calculation shows that the partial derivative is
finite and remains bounded as $z\to 0+$. The other contribution to
the scalar product is finite because of the assumption {\sf
(a2)}.\qed

\begin{lemma}\label{AuxLem4}
\begin{displaymath}
  \exists C_{4}>0\ :\qquad
  \Biggl|\Biggl(
  \,e^{-z|\cdot|}\,|\alpha\!-\!\alpha_0|^{\frac{1}{2}},
  \Bigl[\,|\cdot| P_{\alpha}+P_{\alpha} |\cdot|\,\Bigr]
  \,e^{-z|\cdot|}\,|\alpha\!-\!\alpha_{0}|^{\frac{1}{2}}\,
  \Biggr)\Biggr|\leq \frac{C_{4}}{|z|}
\end{displaymath}
\end{lemma}
\PF Since $x\,e^{-x}\leq e^{-1}$ for any $x>0$, using Schwarz
inequality we infer that the expression is bounded by
\mbox{$e^{-1}|z|^{-1}\,\|\alpha\!-\!\alpha_0\|_1\,
\|P_{\alpha}\|_{HS}$}. \qed
\medskip

Now we are able to complete the proof. Using the elementary
inequality $(I+\lambda P_{\alpha})^{-1}\leq C_5\,(1-\lambda
P_{\alpha})$ valid for small $\lambda$ together with the preceding
lemmas we get for $|z^{-1}|\leq C_1\lambda^{-1}$ and all
sufficiently small $\lambda$:
\begin{displaymath}
  \left|\frac{\partial G}{\partial z}\right|\leq C\lambda\,.
\end{displaymath}
Suppose that $z_{1},z_{2}$ are two solutions of the equation
$z=G(z,\lambda)$. They have to fulfill
\begin{displaymath}
  |z_{1}-z_{2}|=\left|\int_{z_{1}}^{z_{2}}
  \frac{\partial G}{\partial z}dz\right|
  \leq
  \left|\int_{z_{1}}^{z_{2}}
  \left|\frac{\partial G}{\partial z}\right|dz\right|
  \leq
  C\lambda|z_{1}-z_{2}|,
\end{displaymath}
hence the uniqueness is ensured for $\lambda<C^{-1}$.\qed

\subsection{Weak-coupling expansion}
The results of the previous section make it possible to derive a
necessary and sufficient condition for existence of a weakly
coupled state, and also to write an expansion of the bound-state
energy.

\begin{thm}\label{ThmExpansion}
Assume \mbox{\sf (a1)}, \mbox{\sf (a2)} and
$\alpha-\alpha_{0}\not\equiv 0$. Then the operator $H_{\alpha}$
has an eigenvalue $E(\lambda)<\nu_{1}$ for all sufficiently small
$\lambda>0$ if and only if \mbox{$\int_{\Real}(\alpha(x)\!
-\!\alpha_{0})\,dx\leq 0$}. In such a case, the eigenvalue is
unique, simple, and obeys
\begin{eqnarray}\label{expansion}
\lefteqn{\sqrt{\nu_{1}-E(\lambda)} \,=\,
-\frac{\lambda}{2}\,|\chi_{1}(0)|^2
  \int_{\Real}(\alpha(x)\!-\!\alpha_{0})\,dx}\nonumber\\
&&
  -\frac{\lambda^2}{4}\Biggl\{|\chi_{1}(0)|^4
  \int_{{\Real}^2}(\alpha(x)\!-\!\alpha_{0})\,|x-x'|\,
  (\alpha(x')\!-\!\alpha_{0})\,dx\,dx'\nonumber\\
&&
  -|\chi_{1}(0)|^2\sum_{n=2}^{\infty}|\chi_{n}(0)|^2
  \int_{{\Real}^2}(\alpha(x)\!-\!\alpha_{0})\,
  \frac{e^{-\sqrt{\nu_{n}-\nu_{1}}|x-x'|}}{\sqrt{\nu_{n}-\nu_{1}}}\,
  (\alpha(x')\!-\!\alpha_{0})\,dx\,dx'\Biggr\}\nonumber\\
&&
  +{\cal O}(\lambda^3)\,.
\end{eqnarray}
\end{thm}
\PF Using the implicit-function theorem we can check that
(\ref{mainthm}) has a unique solution for small $\lambda$, and
that it is given by~(\ref{expansion}). It remains to prove that
such a solution is strictly positive. This is clearly true for
small enough $\lambda$ if $\int(\alpha(x)\!-\!\alpha_{0})dx<0$. If
the integral is zero, we have to check that the quadratic term is
positive, which can be done by using the Fourier transformation in
the same way as in~\cite[Thm.~4.2]{DE1}.\qed

\setcounter{equation}{0}
\section{Narrow-window coupling}\label{Sec.Scaling Behaviour}

\subsection{Motivation: squeezing the ``leaky" part}

If the effective potential is attractive, $\int(\alpha(x)\!-\!
\alpha_{0})\,dx<0$ the formula~(\ref{expansion}) can be rephrased
as
\begin{equation}\label{ourResult}
  E(\lambda)=\nu_{1}(\alpha_0)-c\lambda^2+{\cal O}(\lambda^3)\ ,\ \quad
  c:=\frac{|\chi_{1}(0)|^4}{4}
     \left(\int_{\Real}(\alpha(x)\!-\!\alpha_{0})\,dx\right)^2.
\end{equation}
Hence the asymptotic behaviour is similar to that
of~\cite[Thm.~1.2.]{BGRS} where a straight Dirichlet strip with a
small protrusion is considered.

However, the supremum norm is not the only mean by which the
perturbation weakness can be controlled. To see that recall the
example~\cite{ESTV} of a double waveguide separated by a Dirichlet
barrier with a window of a width $\ell$. It is very different from
the situation considered above, since the Dirichlet condition
corresponds formally to \mbox{$\alpha_{0}=\infty$}. Nevertheless,
it has a weakly coupled state if the window is narrow, $\ell\ll
d$, where $d:=\max\{d_1,d_2\}$. It was conjectured in ~\cite{ESTV}
that
\begin{equation}\label{windowResult}
  E(\ell)=\left(\frac{\pi}{d}\right)^2-c(\nu)\ell^4+{\cal O}(\ell^5),
\end{equation}
where the parameter $\nu$ describes the waveguide asymmetry,
$$ \nu:=\frac{\min\{d_1,d_2\}}{\max\{d_1,d_2\}}. $$
In~\cite{EV} the conjecture was supported by proving two-sided
bounds by multiples of $\ell^4$ for the energy gap. Recently
Popov~\cite{Po} proved that the formula (\ref{windowResult}) is
valid with
\begin{equation}\label{asympt-coef}
  c(\nu) = \left\lbrace \begin{array}{ccc} \left(2\pi^3\over
  d^3\right)^2 & \quad \dots \quad & \nu=1 \\ \left(\pi^3\over
  d_+^3\right)^2 & \quad \dots \quad & \nu<1 \end{array} \right.
\end{equation}
where $d_+:=\max\{d_1,d_2\}$. Recall that a similar quartic
behaviour is known from waveguides with a {\em critical} local
deformation \cite{EV1} as well as for slightly bent or broken
tubes~\cite{ABGM, DE1} where the leading term in the energy gap is
proportional to the fourth power of the bending angle.

While our method does not allow us to include the case of a
Dirichlet barrier, since such a boundary condition changes the
form domain of the Hamiltonian, it is useful to investigate in our
setting the situation when the weak-coupling limit consists of
squeezing the ``leaky" part while keeping $\|\alpha\!-\!
\alpha_{0}\|_{\infty}$ fixed. We will achieve that by by
introducing a longitudinal scaling of the coupling function,
\begin{equation}
  \alpha_{\sigma}(x):=\alpha\left(\frac{x}{\sigma}\right)\,,
\end{equation}
with the scaling parameter $\sigma\in(0,1]$ and considering the
limit $\sigma\to 0+$.

The argument proceeds in a similar way as above. The main tool is
again Corrolary~\ref{K=-1}(iv) which is not affected by the
scaling. However, as~Remark~\ref{formallimit}(i) below shows, one
cannot apply now the implicit-function theorem to derive an
expansion for $\kappa_1$ analogous to~(\ref{expansion}). To avoid
this difficulty, we employ the simpler decomposition~(\ref{K=L+P})
inspired by~\cite{Si}. It requires a stronger decay, namely
\begin{description}
\item[\mbox{\sf (a2')}] $\:\alpha(\cdot)\!-\!\alpha_{0}\in
     \si(\Real,|x|^2\,dx)$\,.
\end{description}
Let us show how the above results look like in the changed
setting.
\subsection{Modified lemmas}
First note that $\lim_{\sigma\to 0+}k^2=\nu_{1}$, \ie,
$\kappa_{1}\!\to 0$ as $\sigma\to 0+$. We put
\beq\label{K=L+P}
  K_{\alpha_{\sigma}}^k=
  L_{\alpha_{\sigma}}+P_{\alpha_{\sigma}}=
  L_{\alpha_{\sigma}}+M_{\alpha_{\sigma}}+N_{\alpha_{\sigma}}
\eeq
where the kernels are given by
\begin{eqnarray}\label{LMN2}
  L_{\alpha_{\sigma}}(x,x') &=&
                       |\alpha_{\sigma}(x)\!-\!\alpha_{0}|^{\frac{1}{2}}
                       \ \frac{|\chi_{1}(0)|^2}{2\kappa_{1}}
                       \,(\alpha_{\sigma}(x')\!-\!\alpha_{0})^{\frac{1}{2}}
                       \nonumber\\
  M_{\alpha_{\sigma}}(x,x') &=&
                       |\alpha_{\sigma}(x)\!-\!\alpha_{0}|^{\frac{1}{2}}
                       \ \frac{|\chi_{1}(0)|^2}{2\kappa_{1}}
                       \,(e^{-\kappa_{1}|x-x'|}-1)
                       \,(\alpha_{\sigma}(x')\!-\!\alpha_{0})^{\frac{1}{2}}
                       \qquad\\
  N_{\alpha_{\sigma}}(x,x') &=&
                       |\alpha_{\sigma}(x)\!-\!\alpha_{0}|^{\frac{1}{2}}
                       \sum_{n=2}^{\infty}
                       \frac{|\chi_{n}(0)|^2}{2\kappa_{n}}
                       \,e^{-\kappa_{n}|x-x'|}
                       \,(\alpha_{\sigma}(x')\!-\!\alpha_{0})^{\frac{1}{2}}\,.
                       \nonumber
\end{eqnarray}
Next we define
\begin{eqnarray}
  M_{0_{\sigma}}(x,x')
&:=& -|\alpha_{\sigma}(x)\!-\!\alpha_{0}|^{\frac{1}{2}}
  \ \frac{|\chi_{1}(0)|^2}{2}\ |x-x'|
  \ (\alpha_{\sigma}(x')\!-\!\alpha_{0})^{\frac{1}{2}} \nonumber\\
&&\label{M0}\\
  N_{0_{\sigma}}^\beta(x,x')
&:=& |\alpha_{\sigma}(x)\!-\!\alpha_{0}|^{\frac{1}{2}}
  \sum_{n=2}^{\infty}\frac{|\chi_{n}(0)|^2}{2}\
  \frac{e^{-\beta\sqrt{\nu_{n}-\nu_{1}}|x-x'|}}{\beta\sqrt{\nu_{n}-\nu_{1}}}
  \ (\alpha_{\sigma}(x')\!-\!\alpha_{0})^{\frac{1}{2}}\,. \nonumber
\end{eqnarray}
In analogy with Proposition~\ref{HS-lims} we have
\begin{prop}\label{HS-lims2}
Let the assumptions \mbox{\sf (a1)}, \mbox{\sf (a2')} be valid,
then
\begin{equation}
  \lim_{\kappa_{1}\to 0}\|M_{\alpha_{\sigma}}-M_{0_{\sigma}}\|_{HS}=0
  \qquad\mbox{\rm and}\qquad
  \lim_{\kappa_{1}\to 0}\|N_{\alpha_{\sigma}}-N_{0_{\sigma}}^1\|_{HS}=0\,.
\end{equation}
\end{prop}
\PF One has to check the first assertion, because the second one
in Proposition~\ref{HS-lims} does not change. The operator
$M_{0_{\sigma}}$ is Hilbert-Schmidt,
\begin{displaymath}
  \|M_{0}\|_{HS}^2\leq\sigma^4\,\frac{|\chi_{1}(0)|^4}{4}
  \int_{\Real^2}|\alpha(x)\!-\!\alpha_{0}|\ (|x|^2+|x'|^2)\
  |\alpha(x')\!-\!\alpha_{0}|\,dx\,dx'<\infty.
\end{displaymath}
Since $\lim_{\kappa_{1}\to 0}M_{\alpha_{\sigma}}=M_{0_{\sigma}}$
and
$|M_{\alpha_{\sigma}}(x,x';\kappa_{1})|\leq|M_{0_{\sigma}}(x,x')|$,
the result follows by means of the dominated convergence
theorem.\qed \medskip

It is easy to check that the norms of the operators
$M_{\alpha_{\sigma}},N_{\alpha_{\sigma}}$ can be made arbitrarily
small by choosing $\sigma$ small enough, because
\begin{equation}\label{sigma^4}
  \|M_{\alpha_{\sigma}}\|_{HS}^2\leq K_1\,\sigma^4
  \qquad\mbox{and}\qquad
  \|N_{\alpha_{\sigma}}\|_{HS}^2\leq K_2\,\sigma\,.
\end{equation}
Lemma~\ref{boundAN} remains valid without any changes. Since its
first claim is obtained by an algebraic manipulation, it holds for
the operator $M_{\alpha_{\sigma}}$ as well. Finally, it is easy to
see that Lemmas~\ref{AuxLem1}--\ref{AuxLem3} modify as follows:
\begin{lemma}
If $\alpha\!-\!\alpha_{0}\not\equiv 0$, then
$|z|^{-1}\leq C_{1}\sigma^{-1}$
for $\sigma$ small.
\end{lemma}

\begin{lemma}
For sufficiently small $z$,
\begin{displaymath}
  \left\|\frac{\partial P_{\alpha_{\sigma}}}{\partial z}\right\|
  <C_{2}\frac{\sqrt{\sigma}}{|z|}.
\end{displaymath}
\end{lemma}

\begin{lemma}
\begin{displaymath}
  \forall z_{0}\in\Real\quad
  \exists C_{3}>0\quad\forall z\in[0,z_{0}]\ :\
  \left|\left(|\alpha_{\sigma}\!-\!\alpha_{0}|^{\frac{1}{2}},
  \frac{\partial P_{\alpha_\sigma}}{\partial z}\,
  |\alpha_{\sigma}\!-\!\alpha_{0}|^{\frac{1}{2}}\right)\right|
  \leq C_{3}\sigma^2
\end{displaymath}
\end{lemma}
\PF
Here one has only to show that $\partial
M_{\alpha_{\sigma}}/\partial z$ remains bounded as $z\to 0+$ but
this is clear from an elementary inequality, $1-e^{-x}-x
e^{-x}\leq x^2$ for any $x\geq 0$. The rest of the argument
follows the proof of Lemma~\ref{AuxLem3} with the rescaled
integration variables, $(x,x')\mapsto \sigma(x,x')$.\qed

\subsection{The results for the scaled case}

With these preliminaries we can now formulate and prove a
counterpart of Theorem~\ref{mainThm}.
\begin{thm}
Let the assumptions \mbox{\sf (a1)}, \mbox{\sf (a2')} be valid.
Then $H_{\alpha_{\sigma}}$ has for $\sigma$ small enough at most
one simple eigenvalue $E(\sigma)<\nu_{1}$, and this happens if and
only if
\begin{equation}\label{mainthm2}
  \sqrt{\nu_{1}-E}\equiv\kappa_{1}=
  -\frac{|\chi_{1}(0)|^2}{2}
  \left((\alpha_{\sigma}\!-\!\alpha_{0})^{\frac{1}{2}},
  (I+P_{\alpha_{\sigma}})^{-1}
  |\alpha_{\sigma}\!-\!\alpha_{0}|^{\frac{1}{2}}\right)
\end{equation}
has a solution $\kappa_{1}>0$.
\end{thm}
\PF We mimick the proof of Theorem~\ref{mainThm}. Since
$\|P_{\alpha_{\sigma}}\|<1$ holds for small enough $\sigma$, we
may write
\begin{displaymath}
  (I+K_{\alpha_{\sigma}}^k)^{-1}=
  \left[I+(I+P_{\alpha_{\sigma}})^{-1}L_{\alpha_{\sigma}}\right]^{-1}
  (I+P_{\alpha_{\sigma}})^{-1}.
\end{displaymath}
The operator $(I+P_{\alpha_{\sigma}})^{-1}L_{\alpha_{\sigma}}$ has
just one eigenvalue $(\psi,\varphi)$ with
\begin{eqnarray*}
  \psi&:=&\frac{|\chi_1(0)|^2}{2\kappa_1}\
  (\alpha_{\sigma}(\cdot)-\alpha_{0})^{\frac{1}{2}}\\
  \varphi&:=&(I+P_{\alpha_{\sigma}})^{-1}\
  |\alpha_{\sigma}(\cdot)-\alpha_{0}|^{\frac{1}{2}}\,.
\end{eqnarray*}
Putting it equal to $-1$ we get the implicit
equation~(\ref{mainthm2}). As above, we can introduce
\begin{equation}
  G(z,\sigma):=\frac{|\chi_{1}(0)|^2}{2}
                \left(|\alpha_{\sigma}\!-\!\alpha_{0}|^{\frac{1}{2}},
                (I+P_{\alpha_{\sigma}})^{-1}
                |\alpha_{\sigma}\!-\!\alpha_{0}|^{\frac{1}{2}}\right)
\end{equation}
and derive the estimate
\begin{eqnarray*}
  \left|\frac{\partial G}{\partial z}\right|<C\,\sigma^2\,.
\end{eqnarray*}
which means that the uniqueness is now ensured for
$\sigma<C^{-1/2}$.\qed
\medskip

The bound-state criterion and the asymptotic expansion with
respect to the scaling parameter $\sigma$ look now as follows:
\begin{thm}
Let $\alpha$ satisfy the assumptions \mbox{\sf (a1)}, \mbox{\sf
(a2')} and $\alpha\!-\!\alpha_{0}\not\equiv0$. Then
$H_{\alpha_{\sigma}}$ has an eigenvalue $E(\sigma)<\nu_{1}$ for
all $\sigma$ small enough if and only if
\mbox{$\int_{\Real}(\alpha(x)\!-\!\alpha_{0})\,dx\leq 0$}. If this
condition holds, then the eigenvalue is unique, simple, and obeys
\begin{eqnarray}\label{expansion2}
\lefteqn{\sqrt{\nu_{1}-E(\sigma)} \,=\,
-\frac{\sigma}{2}\,|\chi_{1}(0)|^2
  \int_{\Real}(\alpha(x)\!-\!\alpha_{0})\,dx} \nonumber\\
&&\!+\frac{\sigma^2}{4}\,|\chi_{1}(0)|^2
  \sum_{n=2}^{\infty}|\chi_{n}(0)|^2
  \int_{\Real^2}(\alpha(x)\!-\!\alpha_{0})\,
  \frac{e^{-\sigma\sqrt{\nu_{n}-\nu_{1}}|x-x'|}}{\sqrt{\nu_{n}-\nu_{1}}}\,
  (\alpha(x')\!-\!\alpha_{0})\,dx\,dx'\nonumber\\
&&\!+{\cal O}(\sigma^3)\,.
\end{eqnarray}
\end{thm}
\PF Writing
\begin{equation}\label{expanP2}
  (I+P_{\alpha_{\sigma}})^{-1}=
  I-P_{0_{\sigma}}-(P_{\alpha_{\sigma}}-P_{0_{\sigma}})
  +P_{\alpha_{\sigma}}^2(I+P_{\alpha_{\sigma}})^{-1},
\end{equation}
we see that~(\ref{mainthm2}) has a unique solution for $\sigma$
small which is given by~(\ref{expansion2}). It is only important
at that to notice that although $\sigma$ does enter the
expansion~(\ref{expanP2}) explicitly, it appears after
inserting~(\ref{expanP2}) into~(\ref{mainthm2}) because of the
integration. This is also why we know that the last term
in~(\ref{expanP2}) does not contribute to the leading term
in~(\ref{expansion2}).) The rest of the proof concerning the
strict positivity of such a solution proceeds in exactly the same
way as in Theorem~\ref{ThmExpansion}.\qed

\begin{rems}\label{formallimit}{\rm
    {\em (i)} We cannot expand the exponential in the
quadratic term of~(\ref{expansion2}) because the sum may not
converge.\\
    {\em (ii)} Notice that owing to~(\ref{sigma^4}),
the expansion~(\ref{expansion2}) does not contain the term arising
from $M_{0_\sigma}$. This is a substantial difference from the
analogous expansion~(\ref{expansion}).\\
    {\em (iii)} In order to be able to compare the present case with
the pierced Dirichlet barrier mentioned in the opening of this
section, one should perform the limit $\alpha_{0}\to\infty$
assuming that $\alpha\equiv 0$ holds on a small compact. We
observe that $\chi_{n}(0)$ decays like ${\cal O}(\alpha_{0}^{-1})$
as $\alpha_0\to\infty$, so the first term in~(\ref{expansion2})
vanishes after the limit. To get the
behaviour~(\ref{windowResult}) one would need to interchange the
limit with the summation in the next term; it is not clear whether
this can be done. }
\end{rems}

\section{A bound on the number of eigenvalues}
\subsection{A general SKN-type bound}

It is known that while a naive application of the Birman-Schwinger
technique fails to yield an estimate on the bound state number in
dimensions one and two, a simple trick invented independently by
Seto~~\cite{Se}, Klaus~~\cite{Kl}, and Newton~\cite{Ne} does the
job. In this section we apply the idea to the measure-induced
interaction in a strip to get an upper bound for the number of
eigenvalues of our Hamiltonian below the essential spectrum
$\sigma_\mathrm{ess}(H_\alpha)=[\nu_1(\alpha_0),\infty)$. The
argument follows closely the considerations
of~\cite[Sec.~3]{BEKS}, hence we put emphasis again on the
modifications. Hereafter, we assume \mbox{\sf (a1)}, \mbox{\sf
(a2')} because we shall employ the simpler
decomposition~(\ref{K=L+P}) to single out the singularity in the
kernel of of $K_\alpha^k$.

We denote by $\gamma$ the negative part of $\alpha\!-\!\alpha_0$, \ie,
$\gamma:=\max\{0,-(\alpha\!-\!\alpha_0)\}$, and put
\begin{eqnarray*}
  \mu_1(\lambda)
  &:=&
  \inf\{\mathcal{E}_{(\alpha_0-\lambda\gamma) m}(\psi,\psi) \ |\
  \psi\in\sobi(\Omega), \|\psi\|=1\} \\ \\
  \mu_n(\lambda)
  &:=&
  \sup_{\varphi_j\in\sii(\Omega)}
  \inf\Big\{\mathcal{E}_{(\alpha_0-\lambda\gamma) m}(\psi,\psi) \ |\
  \psi\in\sobi(\Omega), \|\psi\|=1, \\
  && \hspace{25ex}
  (\psi,\varphi_j)=0\,, \; j=1,\dots,n\!-\!1
  \Big\}
\end{eqnarray*}
for any $\lambda\in[0,1)$ and all $n\in\Nat\!\setminus\!\{0\}$. We
recall that the measure $\gamma m$ is finite by assumption and
belongs to the generalized Kato class (\cf~Section~\ref{Sec.BS1}).

In analogy with~\cite[Lemma~3.3.]{BEKS} we find that
$\lambda\mapsto\mu_n(\lambda)$ is a non-increasing continuous
function on $[0,1)$ and $\mu_n(0)=\nu_1(\alpha_0)$ for all
$n\in\Nat$. Mimicking further the second part of the proof of
Proposition~\ref{HS-lims} one can show that
$K_{\alpha_0-\lambda\gamma}^{i\kappa}$ is compact for $\kappa$
large enough, since it has a finite Hilbert-Schmidt norm. It is
useful to introduce the standard family of Schatten norms,
$$
  \|K\|_p:=\left(\sum_{j\in J}s_j(K)^p\right)^\frac{1}{p}
$$
for all $1\leq p<\infty$, where $\{s_j(K)\}_{j\in J}$ is the
family of eigenvalues of $|K|$; each eigenvalue is counted
according to its multiplicity as an eigenvalue of $|K|$. We denote
by $N_E$ the number of eigenvalues (counting multiplicity) of
$H_\alpha$ which are smaller than $E$,   and by $\# A$ the
cardinality of the set $A$.

The crux of the BS method is the recognition that the number $N_E$
is equal to the number of eigenvalues of $K_\alpha^{\sqrt{E}}$
that are not less than 1. It immediately follows from the form
version of the minimax principle that $N_E\leq N_E^-$, the number
of eigenvalues (counting multiplicity) of $H_{\alpha_0-\gamma}$
smaller than $E$. In analogy with~\cite[Thm.~3.3.]{BEKS} we
therefore have
\begin{prop} \label{num-est}
$N_E\leq\#\{j\in J\,|\,s_j(K_{\alpha_0-\gamma}^{\sqrt{E}})\geq
1\}$ holds for $E<\nu_1(\alpha_0)$. In particular, we have
$N_E\leq\|K_{\alpha_0-\gamma}^{\sqrt{E}}\|_p^p$ for any $1\leq
p<\infty$.
\end{prop}

This is the naive application mentioned above. It is not
satisfactory in our situation, since the corresponding Green's
function in $K_\alpha^k$ diverges for $k^2\to\nu_1(\alpha_0)$ --
\cf~(\ref{kernel}). The SKN-trick is based on the observation that
this singularity does not depend effectively on the spectral
parameter and corresponds therefore to just one bound state which
can be taken into account separately.
\begin{thm}\label{BS-bound}
Suppose that $\|\gamma\|_1\not=0$. Then the number
$N_{\nu_1(\alpha_0)}$ of eigenvalues (counting multiplicity) below
the threshold of the essential spectrum of $H_\alpha$ satisfies
the bound
\begin{eqnarray*}
  \lefteqn{N_{\nu_1(\alpha_0)}-1\leq} \\
  && \\
  &&
  \phantom{+}
  \frac{|\chi_1(0)|^4}{4\|\gamma\|_1^2}\int_{\Real^4}
  |x_1-x_2|\,\Big(|x_1-x_2| \\
  &&
  \phantom{+}\quad
  +|x_3-x_4|-|x_1-x_3|-|x_2-x_4|\Big)\ \prod_{i=1}^4 \gamma(x_i)\,dx_i \\
  &&
  +\frac{1}{4\|\gamma\|_1^2}\sum_{m,n=2}^\infty
  \frac{|\chi_m(0)\chi_n(0)|^2}{\sqrt{\nu_m-\nu_1}\sqrt{\nu_n-\nu_1}}\int_{\Real^4}
  e^{-\sqrt{\nu_m-\nu_1}|x_1-x_2|} \left(e^{-\sqrt{\nu_n-\nu_1}|x_1-x_2|}\right. \\
  &&
  \phantom{+}\quad\left.
  +e^{-\sqrt{\nu_n-\nu_1}|x_3-x_4|}
  -e^{-\sqrt{\nu_n-\nu_1}|x_1-x_3|}-e^{-\sqrt{\nu_n-\nu_1}|x_2-x_4|}\right)
  \ \prod_{i=1}^4 \gamma(x_i)\,dx_i \\
  &&
  -\frac{|\chi_1(0)|^2}{2\|\gamma\|_1^2}
  \sum_{n=2}^\infty\frac{|\chi_n(0)|^2}{\sqrt{\nu_n-\nu_1}}
  \int_{\Real^4} |x_1-x_2| \left(e^{-\sqrt{\nu_n-\nu_1}|x_1-x_2|}\right. \\
  &&
  \phantom{+}\quad\left.
  +e^{-\sqrt{\nu_n-\nu_1}|x_3-x_4|}
  -e^{-\sqrt{\nu_n-\nu_1}|x_1-x_3|}-e^{-\sqrt{\nu_n-\nu_1}|x_2-x_4|}\right)
  \ \prod_{i=1}^4 \gamma(x_i)\,dx_i\,.
\end{eqnarray*}
\end{thm}
\PF It is an obvious modification of the proofs in~\cite{Ne}.
Borrowing the notation from this article and taking
~(\ref{K=L+P}), (\ref{LMN2}) into account, we can write
$$
  K_{\alpha_0-\gamma}=\xi(\varphi,\cdot)\varphi+P_{\alpha_0-\gamma}
$$
with $\xi = -\frac{|\chi_1(0)|^2}{2\kappa_1}\|\gamma\|_1$,
$\varphi= \gamma^{1/2}/\|\gamma\|_1^{-1/2}$, and
$P_{\alpha_0-\gamma}=M_{\alpha_0-\gamma}+N_{\alpha_0-\gamma}\,$.
We use the inequality obtained in~\cite[pp.~123]{Ne} for
$\xi\to\infty$,
$$
  N_{\nu_1(\alpha_0)}^-\leq 1+\tr P_0^2-2(\varphi,P_0^2\varphi)
  +(\varphi,P_0\varphi)^2,
$$
and substitute $P_0=M_0+N_0^1$, where $M_0,N_0^1$ are given
by~(\ref{M0}); this leads to the
desired result.\qed

\subsection{A ``rectangular well''  example}
To illustrate the above result let us apply it to the example
analyzed in~\cite[Sec.~4]{EK1} in which $\alpha$ is a steplike
function:
$$
  \alpha(x):=\left\{
  \begin{array}{lcl}
    \alpha_1 & \quad \mathrm{if} & \ |x|<a \\
    && \\
    \alpha_0 & \quad \mathrm{if} & \ |x|\geq a
  \end{array}\right.
$$
for some real $\alpha_1<\alpha_0$. Under the last condition the
waveguide has a nontrivial discrete spectrum. Since
$\gamma:=\alpha_0\!-\!\alpha_1$ is a constant on its support, we
can evaluate the integrals of Theorem~\ref{BS-bound} obtaining
\begin{eqnarray}\label{BS-bound-ex}
  \lefteqn{N_{\nu_1(\alpha_0)}\leq
  1+\frac{8}{45}|\chi_1(0)|^4 \gamma^2 a^4} \nonumber\\
  &&
  +\ \frac{\gamma^2 a^4}{2} \sum_{m,n=2}^\infty
  \frac{|\chi_m(0)\chi_n(0)|^2}{\tilde\kappa_m\tilde\kappa_n}
  \Bigg\{
  \frac{2}{a(\tilde\kappa_m+\tilde\kappa_n)}
  +\frac{4}{a^2\tilde\kappa_m\tilde\kappa_n}
  -\frac{1-e^{-2a(\tilde\kappa_m+\tilde\kappa_n)}}
  {a^2(\tilde\kappa_m+\tilde\kappa_n)^2} \nonumber\\
  && \quad
  +\frac{(1-e^{-2a\tilde\kappa_m})(1-e^{-2a\tilde\kappa_n})}
  {a^3\tilde\kappa_m\tilde\kappa_n(\tilde\kappa_m+\tilde\kappa_n)}
  -\frac{4\left[\tilde\kappa_m(1-e^{-2a\tilde\kappa_n})
  +\tilde\kappa_n(1-e^{-2a\tilde\kappa_m})
  \right]}
  {a^3(\tilde\kappa_m\tilde\kappa_n)^2}
  \nonumber\\
  && \quad
  +\frac{2\left[\tilde\kappa_m^3(1-e^{-2a\tilde\kappa_n})
  -\tilde\kappa_n^3(1-e^{-2a\tilde\kappa_m})\right]}
  {a^3(\tilde\kappa_m\tilde\kappa_n)^2
  (\tilde\kappa_m^2-\tilde\kappa_n^2)}
   +\frac{2(1-e^{-2a\tilde\kappa_m})(1-e^{-2a\tilde\kappa_n})}
  {a^4\tilde\kappa_m^2\tilde\kappa_n^2}
  \Biggr\} \nonumber\\
  &&
  -\ 2|\chi_1(0)|^2 \gamma^2 a^3
  \sum_{n=2}^\infty\frac{|\chi_n(0)|^2}{\tilde\kappa_n}
  \Biggl(-\frac{2}{3a\tilde\kappa_n}+\frac{2}{a^2\tilde\kappa_n^2}
  -\frac{1-e^{-2a\tilde\kappa_n}}{3a^2\tilde\kappa_n^2}
  -\frac{2}{a^3\tilde\kappa_n^3} \nonumber\\
  && \quad
  +\frac{1-e^{-2a\tilde\kappa_n}}{a^4\tilde\kappa_n^4} \Biggr)
\end{eqnarray}
where $\tilde\kappa_n$ abbreviates $\sqrt{\nu_n-\nu_1}$.

To assess this bound, compare it with the one following from a
simple bracketing argument and the minimax principle
~\cite[Sec.~4.1.]{EK1} which reads
\begin{equation} \label{brack}
  N_{\nu_1(\alpha_0)}\leq
  1+\left[ \frac{2a}{\pi}\sqrt{\nu_1(\alpha_0)-\nu_1(\alpha_1)}
  \right]\,,
\end{equation}
where $[\cdot]$ denotes the entire part. The \rhs\/ is a
``linearly increasing" step function with respect to the wi ndow
halfwidth $a$. In distinction to (\ref{BS-bound-ex}), however, the
bracketing argument yields in this example also a tight lower
bound which differ just by one from (\ref{brack}). The bound
(\ref{BS-bound-ex}) is not only more complicated, but it increases
much faster with $a$; the comparison illustrates once more that
while the Birman-Schwinger method is efficient for weak coupling,
it may provide results far from optimal for strongly coupled
systems.

\bigskip
\noindent\textbf{Acknowledgments.} We benefited from a
discussion with W.~Renger as well as from useful remarks of the
referee. The work has been partially supported by the GA AS
Grant 1048801.

\newpage

\end{document}